\documentclass[12pt]{article}
\usepackage{graphicx}

\usepackage{scicite}

\usepackage{times}

\newenvironment{sciabstract}{%
\begin{quote} \bf}
{\end{quote}}

\newcounter{lastnote}

\begin{document}

\title{The Detection of a Population of Submillimeter$-$Bright,
  Strongly-Lensed Galaxies}

\author{
  Mattia Negrello$^{1\ast}$, R. Hopwood$^{1}$, G. De Zotti$^{2,3}$, A. Cooray$^{4}$, A. Verma$^{5}$,  \\
  J. Bock$^{6,7}$, D. T. Frayer$^{8}$, M. A. Gurwell$^{9}$, A. Omont$^{10}$, R. Neri$^{11}$, \\
  H. Dannerbauer$^{12}$,  L. L. Leeuw$^{13,14}$, E. Barton$^{4}$,  J. Cooke$^{4,7}$, S. Kim$^{4}$, \\
  E. da Cunha$^{15}$, G. Rodighiero$^{16}$, P. Cox$^{11}$, D. G. Bonfield$^{17}$, M. J. Jarvis$^{17}$, \\
  S. Serjeant$^{1}$, R.~J. Ivison$^{18,19}$, S. Dye$^{20}$, I. Aretxaga$^{21}$, D.~H. Hughes$^{21}$, E.~Ibar$^{18}$, \\
  F. Bertoldi$^{22}$, I. Valtchanov$^{23}$, S. Eales$^{20}$,  L. Dunne$^{24}$, S. P. Driver$^{25}$, \\
  R. Auld$^{20}$, S. Buttiglione$^{2}$, A. Cava$^{26,27}$, C.~A. Grady$^{28,29}$, \\
  D. L. Clements$^{30}$, A. Dariush$^{20}$, J. Fritz$^{31}$, D. Hill$^{25}$, \\
  J.~B. Hornbeck$^{32}$,  L. Kelvin$^{25}$, G. Lagache$^{33,34}$,  M. Lopez-Caniego$^{35}$, \\
  J. Gonzalez-Nuevo$^{3}$,  S. Maddox$^{24}$, E. Pascale$^{20}$,  M. Pohlen$^{20}$,  E. E. Rigby$^{24}$, \\
  A. Robotham$^{25}$,  C. Simpson$^{36}$,   D. J. B. Smith$^{24}$,   P. Temi$^{37}$,   M. A. Thompson$^{17}$, \\
  B. E. Woodgate$^{38}$,  D. G. York$^{39}$,  J. E. Aguirre$^{40}$,   A. Beelen$^{34}$,  A. Blain$^{7}$, \\
  A. J. Baker$^{41}$,  M. Birkinshaw$^{42}$, R. Blundell$^{9}$,  C. M. Bradford$^{6,7}$,  D. Burgarella$^{43}$, \\
  L. Danese$^{3}$,  J. S. Dunlop$^{18}$,  S. Fleuren$^{44}$,  J. Glenn$^{45}$, A. I. Harris$^{46}$,  \\
  J. Kamenetzky$^{45}$, R. E. Lupu$^{40}$,   R. J. Maddalena$^{8}$,  B. F. Madore$^{47}$, \\
  P. R. Maloney$^{45}$, H. Matsuhara$^{48}$, M.~J. Micha{\l}owski$^{19}$, E. J. Murphy$^{49}$,  \\
  B. J. Naylor$^{6}$,   H. Nguyen$^{6}$,  C. Popescu$^{50}$,   S. Rawlings$^{5}$,  \\
  D. Rigopoulou$^{5,51}$,  D. Scott$^{52}$,  K. S. Scott$^{40}$,  M. Seibert$^{47}$,  I. Smail$^{53}$, \\
  R. J. Tuffs$^{54}$, J. D. Vieira$^{7}$, P. P. van der
  Werf$^{19,55}$, J. Zmuidzinas$^{6,7}$ }


\date{}

\maketitle
$~$ \\
$^{1}$Department of Physics and Astronomy, The Open University, Walton Hall, Milton Keynes, MK7 6AA, UK \\
$^{2}$Istitute Nazionale di Astrofisica (INAF), Osservatorio Astronomico di Padova, Vicolo Osservatorio 5, I-35122 Padova, Italy \\
$^{3}$Scuola Internazionale Superiore di Studi Avanzati (SISSA), Via Bonomea 265, I-34136 Trieste, Italy \\
$^{4}$Department of Physics and Astronomy, University of California, Irvine, CA 92697, USA \\
$^{5}$Oxford Astrophysics, Denys Wilkinson Building, University of Oxford, Keble Road, Oxford, OX1 3RH,  uk \\
$^{6}$Jet Propulsion Laboratory (JPL), Pasadena CA, 91009, USA \\
$^{7}$California Institute of Technology, Pasadena, CA 91125, USA \\
$^{8}$National Radio Astronomy Observatory Post Office Box 2, Green Bank, WV 24944, USA \\
$^{9}$Harvard-Smithsonian Center for Astrophysics, Cambridge, MA 02138, USA \\
$^{10}$Institut d'Astrophysique de Paris, Universitte Pierre et Marie Curie and CNRS, 98 bis boulevard Arago, 75014 Paris, France \\
$^{11}$Institut de Radioastronomie Millim{\'e}trique (IRAM), 300 rue de la piscine, 38406 Saint-Martin d'H{\`e}res, France \\
$^{12}$Laboratoire Astrophysique, Instrumentation et Mod$\rm\acute{e}$lisation Paris Saclay, Commissariat $\rm\grave{a}$l'$\acute{E}$nergie Atomique (CEA)/Direction des Sciences de la Mati$\rm\grave{e}$re (DSM) - CNRS - Universit$\rm\acute{e}$ Paris Diderot, Institut de recherche sur les lois fondamentales de l'Univers (Irfu)/Service d'Astrophysique, CEA Saclay, Orme des Merisiers, F-91191 Gif-sur-Yvette Cedex, France \\
$^{13}$ Physics Department, University of Johannesburg, Post Office Box 524, Auckland Park 2006, South Africa \\
$^{14}$SETI Institute, 515 North Whisman Avenue Mountain View CA, 94043, USA \\
$^{15}$Department of Physics, University of Crete, 71003 Heraklion, Greece \\
$^{16}$Dipartimento di Astronomia, Universit$\rm\acute{a}$ di Padova, Vicolo Osservatorio 2, I-35122 Padova, Italy \\
$^{17}$Centre for Astrophysics Research, Science and Technology Research Institute, University of Hertfordshire, Herts AL10 9AB, UK \\
$^{18}$UK Astronomy Technology Center, Royal Observatory Edinburgh, Edinburgh, EH9 3HJ, UK \\
$^{19}$ Scottish Universities Physics Alliance, Institute for Astronomy, University of Edinburgh, Royal Observatory,Edinburgh, EH9 3HJ, UK \\
$^{20}$School of Physics and Astronomy, Cardiff University, The Parade, Cardiff, CF24 3AA, UK \\
$^{21}$Instituto Nacional de Astrof\'{i}sica, \'{O}ptica y Electr\'{o}nica, Apartado Postal 51 y 216, 72000 Puebla, Mexico \\
$^{22}$Argelander Institut fur Astronomie, Universit Bonn, Auf dem H$\rm\ddot{u}$gel 71, 53121 Bonn, Germany \\
$^{23}$Herschel Science Centre, European Space Astronomy Centre, European Space Agency (ESA), Post Office Box 78, 28691 Villanueva de la Ca$\rm\tilde{n}$ada, Madrid, Spain \\
$^{24}$School of Physics and Astronomy, University of Nottingham, University Park, Nottingham NG7 2RD, UK \\
$^{25}$Scottish Universities Physics Alliance, School of Physics and Astronomy, University of St. Andrews, North Haugh, St. Andrews, KY16 9SS, UK \\
$^{26}$Instituto de Astrof\'isica de Canarias, C/V\'ia L\'{a}ctea s/n, E-38200 La Laguna, Spain \\
$^{27}$Departamento de Astrof{\'\i}sica, Universidad de La Laguna (ULL), E-38205 La Laguna, Tenerife, Spain \\
$^{28}$Eureka Scientific, 2452 Delmer Street, Suite 100, Oakland CA 94602, USA \\
$^{29}$Goddard Space Flight Center, Code 667, Greenbelt Road, Greenbelt, MD 20771, USA \\
$^{30}$Astrophysics Group, Physics Department, Blackett Lab, Imperial College London, Prince Consort Road, London SW7 2AZ, UK \\
$^{31}$Sterrenkundig Observatorium, Universiteit Gent, Krijgslaan 281 S9, B-9000 Gent, Belgium \\
$^{32}$Department of Physics and Astronomy, University of Louisville, Louisville, KY 40292, USA \\
$^{33}$Institut d'Astrophysique Spatiale (IAS), B$\rm\hat{a}$timent 121, F-91405 Orsay, France \\
$^{34}$Universit$\rm\acute{e}$ Paris-Sud 11 and CNRS (UMR 8617), 91400 Orsay, France \\
$^{35}$Instituto de Fisica de Cantabria, Consejo Superior de Investigaciones Cient$\rm\acute{i}$ficas - Universidad de Cantabria, Avenue de Los Castros s/n, Santander, 39005, Spain \\
$^{36}$Astrophysics Research Institute, Liverpool John Moores University Twelve Quays House, Egerton Wharf, Birkenhead, CH41 1LD, UK \\
$^{37}$Astrophysics Branch, NASA Ames Research Center, Mail Stop 245-6, Moffett Field, CA 94035, USA \\
$^{38}$NASA Goddard Space Flight Center, Code 667, Greenbelt Road, Greenbelt, MD 20771, USA \\
$^{39}$Department of Astrophysics and The Enrico Fermi Institute, University of Chicago, 5640 South Ellis Avenue, Chicago, IL 60637, USA \\
$^{40}$Department of Physics and Astronomy, University of Pennsylvania, Philadephia, PA 19104, US \\
$^{41}$Rutgers, the State University of New Jersey, Department of Physics and Astronomy, 136 Frelinghuysen Road, Piscataway, NJ 08854-8019, USA \\
$^{42}$HH Wills Physics Laboratory, University of Bristol, Tyndall Avenue, Bristol BS8 1TL, UK \\
$^{43}$Laboratoire d'Astrophysique de Marseille and Aix-Marseille Universit$\rm\acute{e}$, UMR6110 CNRS, 38 rue F. Joliot-Curie, F-13388 Marseille, France \\
$^{44}$School of Mathematical Sciences, Queen Mary, University of London, Mile End Road, London, E1 4NS, UK \\
$^{45}$University of Colorado, Center for Astrophysics and Space Astronomy, 389-UCB, Boulder, CO 80303, USA \\
$^{46}$Department of Astronomy, University of Maryland, College Park, MD 20742 USA \\
$^{47}$Observatories of the Carnegie Institution, 813 Santa Barbara Street, Pasadena, CA 91101, USA \\
$^{48}$Institute for Space and Astronautical Science, Japan Aerospace Exploration Agency, 3-1-1 Yoshinodai, Chuo-ku, Sagamihara 252-5210 Japan \\
$^{49}$Infrared Processing and Analysis Center, Pasadena, CA 91125, USA \\
$^{50}$Jeremiah Horrocks Institute, University of Central Lancashire, Preston PR1 2HE, UK \\
$^{51}$Space Science and Technology Department, Rutherford Appleton Laboratory, Chilton, Didcot, Oxfordshire OX11 0QX, UK \\
$^{52}$University of British Colombia, 6224 Agricultural Road, Vancouver, BC V6T 1Z1, Canada \\
$^{53}$Institute for Computational Cosmology, Durham University, Durham DH1 3LE, UK \\
$^{54}$Max Planck Institut fuer Kernphysik (MPIK), Saupfercheckweg 1, 69117 Heidelberg, Germany \\
$^{55}$Leiden Observatory, Leiden University, Post Office Box 9513, NL - 2300 RA Leiden, The Netherlands \\
$~$\\
$^\ast$ to whom correspondence should be addressed E-mail:
m.negrello@open.ac.uk



\baselineskip24pt




\begin{sciabstract}
  Gravitational lensing is a powerful astrophysical and cosmological
  probe and is particularly valuable at submillimeter wavelengths for
  the study of the statistical and individual properties of dusty
  starforming galaxies.  However the identification of gravitational
  lenses is often time-intensive, involving the sifting of large
  volumes of imaging or spectroscopic data to find few candidates.  We
  used early data from the Herschel Astrophysical Terahertz Large Area
  Survey to demonstrate that wide-area submillimeter surveys can
  simply and easily detect strong gravitational lensing events, with
  close to 100$\%$ efficiency.
\end{sciabstract}

When the light from a distant galaxy is deflected by a foreground mass
$-$ commonly a massive elliptical galaxy or galaxy cluster or group
$-$ its angular size and brightness are increased, and multiple images
of the same source may form. This phenomenon is commonly known as
gravitational lensing \cite{notes1} and can be exploited in the study
of high-redshift galaxy structures down to scales difficult (if not
impossible) to probe with the largest telescopes at present
\cite{Marshall07,Stark08,Swinbank10} and to detect intrinsically faint
objects. Surveys conducted at submillimeter wavelengths can
particularly benefit from gravitational lensing because submillimeter
telescopes have limited spatial resolution and consequently high
source confusion, which makes it difficult to directly probe the
populations responsible for the bulk of background submillimeter
emission \cite{Smail97,Blain99}. In addition, galaxies detected in
blank-field submillimeter surveys generally suffer severe dust
obscuration and are therefore challenging to detect and study at
optical and near infra-red (NIR) wavelengths. By alleviating the
photon starvation, gravitational lensing facilitates follow-up
observations of galaxies obscured by dust and in particular the
determination of their redshift \cite{RR91}.  Previous submillimeter
searches for highly magnified background galaxies have predominantly
targeted galaxy cluster fields \cite{Smail02}.  In fact, a blind
search for submillimeter lensing events requires large area because of
their rarity, and sub-arcseconds angular resolutions to reveal
multiple images of the same background galaxies.  Although the first
requirement has recently been fulfilled, thanks to the advent of the
South Pole Telescope (SPT) \cite{Vieira10} and the Herschel Space
Observatory (Herschel) \cite{Pilbratt10}, the second is still the
prerogative of ground-based interferometric facilities, such as the
Submillimeter Array (SMA) and the IRAM Plateau de Bure Interferometer
(PdBI), which because of their small instantaneous field of view are
aimed at follow-up observations rather than large-area survey
campaigns.  Nevertheless, several authors
\cite{Blain96,Perrotta02,Perrotta03,Negrello07} have suggested that a
simple selection in flux density, rather than surveys for
multiply-imaged sources, can be used to easily and efficiently select
samples of strongly gravitationally-lensed galaxies in wide-area
submillimeter and millimeter surveys.  The explanation for this lies
in the steepness of the number counts (the number of galaxies at a
given brightness) of dust-obscured star-forming galaxies, which are
usually referred to as submillimeter galaxies (SMGs)
\cite{Coppin06}. Because of that, even a small number of
highly-magnified SMGs can substantially affect the shape of the bright
end of the submillimeter source counts enhancing the number of SMGs
seen at bright flux densities than would be expected on the basis of
our knowledge of the un-lensed SMG population (Fig.1).  Furthermore,
the frequency of lensing events is relatively high in the
submillimeter \cite{Blain96} because SMGs are typically at high
redshift ($z>\sim1$) \cite{Chapman05}, and this increases the
probability that a SMG is in alignment with, and therefore lensed by,
a foreground galaxy.  Other important contributors to the bright tail
of the submillimeter counts are low-redshift ($z\le0.1$) spiral and
starburst galaxies \cite{Serjeant05} and higher redshift radio-bright
Active Galactic Nuclei (AGNs) \cite{DeZotti05}; however both of these
are easily identified, and therefore removed, in relatively shallow
optical and radio surveys. Therefore, flux-density limited
submillimeter surveys could provide a sample of lens candidates from
which contaminants can be readily removed, leaving a high fraction
(close to 100$\,\%$) of gravitational lens systems (Fig.$\,$1).
Because this selection of lens candidates relies only on the
properties of the background source (its flux density), it can probe a
wide range of lens properties (such as redshifts and masses) and thus
provide a valuable sample for studying the elliptical properties of
lensing galaxies \cite{Koopmans06} as well as investigating the
detailed properties of the lensed SMGs.

\paragraph*{The submillimeter lens candidate selection at work.}

Although the approach presented above may be more efficient and vastly
more time-effective than those exploited so far in the radio
\cite{Browne01} or the optical \cite{Oguri06,Faure08}, at least
several tens of square degrees (deg$^{2}$) of the sky must be observed
in the submillimeter to produce a statistically significant sample of
strongly lensed objects and a minimal contamination from unlensed
galaxies. This is because the surface density of lensed submillimeter
galaxies is predicted to be lower than $\sim$0.5 deg$^{-2}$, for flux
densities above 100$\,$mJy at 500$\,\mu$m (Fig.$\,$1).  Submillimeter
surveys conducted before the advent of Herschel were either limited to
small areas of the sky \cite{Coppin06,Weiss09}, or were severely
affected by source confusion due to poor spatial
resolution\cite{Devlin09}.  Therefore no previous test of this
selection method has been performed, although the SPT has recently
mapped an area of more than 80 deg$^{2}$ at millimetre wavelengths
\cite{Vieira10} and found an
``excess'' of sources that could be accounted for by a population of gravitationally-lensed objects. \\
The Herschel Astrophysical Terahertz Large Area Survey (H-ATLAS)
\cite{Eales10} represents the largest-area submillimeter survey being
currently undertaken by Herschel. H-ATLAS uses the Spectral and
Photometric Imaging REceiver (SPIRE) \cite{Griffin10,Pascale10} and
the Photodetector Array Camera and Spectrometer (PACS)
\cite{Poglitsch10,Ibar10} instruments and, when completed, will cover
$\sim$550$\,$deg$^{2}$ of the sky from 100 to 500$\,\mu$m. H-ATLAS has
been designed to observe areas of the sky with previously existing
multi-wavelength data: Galaxy Evolution Explorer (GALEX) ultra violet
(UV) data, Sloan Digital Sky Survey (SDSS) optical imaging and
spectroscopy, NIR data from the UK Infrared Telescope (UKIRT) Infrared
Deep Sky Survey (UKIDSS) Large Area Survey (LAS), spectra from the
Galaxy And Mass Assembly GAMA \cite{Driver10} project, radio imaging
data from the Faint Images of the Radio Sky at Twenty-cm (FIRST)
survey and the NRAO Very Large Array Sky Survey (NVSS).  The first
14.4$\,$deg$^{2}$ of the survey, centred on J2000 RA 09:05:30.0 DEC
00:30:00.0 and covering $\sim$3$\%$ of the total area, was observed in
November 2009 as part of the Herschel Science Demonstration Phase
(SDP).  The results were a catalog of $\sim$6600 sources
\cite{Rigby10} with a significance $>5\,\sigma$, in at least one
SPIRE waveband, where the noise, $\sigma$, includes both instrumental
and source confusion noise and corresponds to $\sim$7 to 9 mJy/beam. \\
The Herschel/SPIRE 500$\,\mu$m channel is favourable for selecting
lens candidates, because the submillimeter source counts steepen at
longer wavelengths \cite{Devlin09, Clements10}.  We used theoretical
predictions \cite{Negrello07} to calculate the optimal limiting flux
density, above which it is straightforward to remove contaminants from
the parent sample and maximize the number of strongly lensed
high-redshift galaxies.  The surface-density of un-lensed SMGs is
predicted to reach zero by $S_{\rm500}\sim100\,$mJy \cite{Negrello07}
and these objects are only detectable above this threshold if
gravitationally lensed by a foreground galaxy (Fig.$\,$1).  The
H-ATLAS SDP catalog contains 11 sources with 500$\,\mu$m flux density
above 100$\,$mJy.  Ancillary data in the field revealed that six of
these objects are contaminants; four are spiral galaxies with
spectroscopic redshifts in the range of 0.01 to 0.05 [see
\cite{Baes10} for a detailed analysis of one of these sources], one is
an extended Galactic star forming region, and one is a previously
known radio$-$bright AGN \cite{Gonzalez-Nuevo10}. Although the number
of these sources are few at bright flux densities, the measured
surface densities are consistent with expectations (Fig.$\,$1)
\cite{Serjeant05,DeZotti05}. Exclusion of these contaminants left the
five objects that form our sample of lens candidates (table$\,$S1)
\cite{SOM}, identified as ID9, ID11, ID17, ID81 and ID130.

\paragraph*{Unveiling the nature of the lens candidates.}

For gravitational lensing systems selected at submillimeter
wavelengths, we would expect the lensing galaxy to be seen in optical
and/or NIR images, in which the emission from the lens dominates over the
higher redshift background SMG.  In line with these expectations, all
of the lens candidates have a close counterpart in SDSS or UKIDSS
images (or both).  A likelihood ratio analysis \cite{Smith10} showed
that the probability of a random association between these bright
submillimeter sources and the close optical/NIR counterparts is less
than a few per cent. Therefore the optical and submillimeter emissions
must be physically related, either because they occur within the same
object or because of the effects of gravitational lensing, boosting
the flux of the background source and indirectly affecting the
likelihood ratio calculations. The redshift measurements support the
later scenario. Although the optical/NIR
photometric/spectroscopic redshifts lie in the range of $z\sim0.3$ to 0.9 (table 1 and figs S3 and S4) \cite{SOM}, the redshifts estimated from
the submillimeter/millimeter spectral energy distributions (SED)
(table$\,$2) [following the method described in
\cite{Hughes02,Aretxaga03}], are distinctly different (Table 1). The
lensed SMG photometric redshifts have been confirmed and made more
precise through the spectroscopic detection, in these objects, of
carbon monoxide (CO) rotational line emission which are tracers of
molecular gas assocatiated to star forming environments. Until
recently, these kind of detections were difficult to achieve without
prior knowledge of the source redshift, which required extensive
optical/NIR/radio follow-up observations. Because of the development of
wide-bandwidth radio spectrometers capable of detecting CO lines over
a wide range of redshifts, it is now possible for blind redshift
measurements of SMGs to be taken without relying on optical or NIR
spectroscopy \cite{Weib09,Daddi09}.  ID81 was observed with the Z-Spec
spectrometer \cite{Naylor03,Bradford09} on the California Institute of Technology Submillimeter
Observatory (CSO). The data revealed several CO lines redshifted into the
frequency range of 187 to 310$\,$GHz; the strongest of these lines has
been interpreted as the CO J=7$-$6 line, with an estimated redshift of
$z=3.04$ \cite{Lupu10}. This represents the first blind redshift
determination by means of Z-Spec. We followed up this observation with the PdBI
and detected CO J=3$-$2 and CO J=5$-$4 emission lines, redshifted to
$z=3.042$, confirming the Z-Spec measured redshift \cite{SOM}.  We
also used the Zpectrometer instrument \cite{Harris07,notes2} on the
NRAO Robert C. Byrd Green Bank Telescope (GBT) to obtain an
independent confirmation of the redshift of ID81 (table$\,$1 and fig.$\,$S1) \cite{Frayer10,SOM} and to measure the redshift of ID130.
We detected redshifted CO J=1$-$0 emission at $z=2.625$ in the
spectrum of ID130 (fig.$\,$S1) \cite{Frayer10,SOM}].  This
redshift was confirmed by the PdBI with the observation of CO J=3-2
and CO J=5-4 lines, yielding a redshift of $z=2.626$ \cite{SOM}.  The
Z-Spec spectrometer observed the remaining three lens candidates
\cite{Lupu10} and detected CO lines at redshifts of $z=1.577$ and
$z=1.786$ for ID9 (fig.$\,$S2) \cite{SOM} and ID11, respectively, which
are higher and inconsistent with the redshifts derived from the
optical photometry/spectroscopy (Table$\,$1). The Z-Spec CO
measurements for ID17 are indicative of two redshifts; one, $z=0.942$,
that is in agreement with the optical redshift and a higher one,
$z=2.31$, which is indicative of a more distant
galaxy. \\
To determine the morphological type of the foreground galaxies we
obtained high resolution optical images for all five objects with the
Keck telescope at $g$- and $i$-bands \cite{SOM}. ID9, ID11, ID81 and ID130 all have optical profiles that are consistent
with elliptical galaxies (figs S5 and S6 and table S4) \cite{SOM}. 
The interpretation of the results for ID17 is complicated by the presence of
two partially superimposed galaxies in the optical images (fig.$\,$S7) \cite{SOM}, neither
exhibiting the disturbed morphology expected for lensed objects.  
This indicates that ID17 may be a gravitational lens system
with two foreground lensing masses at similar redshifts ($z\sim0.8$ to 0.9) $-$ possibly a merging system $-$ with some molecular gas
responsible for the CO emission detected by Z-Spec at $z \sim 0.9$ and
confirmed with optical spectroscopy (table 1).  A fit to the
UV/optical/NIR SEDs of ID9, ID11, ID81 and ID130 \cite{notes5}, using
the models of \cite{DaCunha08}, gives stellar masses in
the range of $4 \times 10^{10}$ to $15\times{\rm M}_{\odot}$
(Table 2) and almost negligible
present-day star formation, which is consistent with elliptical galaxies (fig.$\,$2). \\
For all five lens systems the background source appears to be
undetected in the Keck $g$- and $i$-band images, despite the flux
magnification due to lensing. After subtracting the best fit light profile from each lens we found no structure that could be associated with the background source in the residual images (figs$\,$S5 and
S6) \cite{SOM}. We derived 3-$\sigma$ upper limits from the residual maps
(table$\,$S4) \cite{SOM} and corresponding NIR limits from the
UKIDSS images. These upper limits were used to fit the SEDs of the
background sources assuming the models of \cite{DaCunha08}, calibrated to reproduce the UV-to-infrared SEDs of
local, purely star-forming ultra luminous infrared galaxies (ULIRG;
10$^{12}\le$L$_{IR}$/L$_{\odot}<$10$^{13}$) \cite{daCunha10} (fig.$\,$2).  A visual extinction
\cite{notes4} of$ A_{V}$ $>$2 is required to be consistent with the
optical/NIR upper limits (fig.$\,$2 and table$\,$2), confirming
severe dust obscuration in these galaxies along the line-of-sight.
Our results indicate that these submillimeter bright gravitationally lensed galaxies 
would have been entirely missed by standard optical methods of selection. \\
We obtained observations at the SMA for
ID81 and ID130 at 880$\,\mu$m, with the aim of detecting the lensed
morphology of the background galaxy \cite{SOM}. The
SMA images reveal extended submillimeter emission distributed around the
cores of the foreground elliptical galaxies, with multiple peaks (four
main peaks in ID81 and two in ID130), which is consistent with a lensing interpretation of these structures (Fig. 3). The position of these peaks can be used to
directly constrain the Einstein radius $-$ the radius of the
circular region on the sky (the Einstein ring) into which an
extended source would be lensed if a foreground galaxy were exactly
along the line of sight of the observer to the source (for a perfectly circular lens). The Einstein
radius is a measure of the projected mass of the lens, so it can be used to derive the total
(dark plus luminous) mass of the galaxy within the Einstein radius
(table$\,$2) \cite{SOM}. Another measure of the
total mass of a lens is the line-of-sight stellar velocity
dispersion, $\sigma_{\rm v}$. We have estimated $\sigma_{\rm v}$ from
the local Faber$-$Jackson (FJ) relation \cite{Faber76} between
$\sigma_{\rm v}$ and the rest-frame B-band luminosity for elliptical
galaxies. Assuming passive stellar evolution for the lens galaxies, 
which is appropriate for elliptical galaxies, we have
extrapolated their rest-frame K-band luminosity  to $z=0$ [using the
evolutionary tracks of \cite{Willott03}], and then converted this to B-band
luminosity using the B$\,$-$\,$K$\,$=$\,$4.43 color relation from \cite{Huang98}. 
The result was then applied  to the
FJ relation from \cite{Gudehus91}.  Given a mass model for the lens
\cite{SOM}, we can predict the Einstein radius of
the galaxy from the value of $\sigma_{\rm v}$ expected from the FJ relation and compare it with that 
directly measured from the SMA images (Table 2). Although the
value of the Einstein radius derived from the line-of-sight stellar
velocity dispersion is affected by large uncertainties (as a result of
the scatter in the FJ relation)
it is consistent with the value measured in the SMA
images for both ID81 and ID130.  
In order to test whether the properties of the lensing galaxies in our sample are consistent with those of other known lens ellipticals at similar redshift, we compared the V-band mass-to-light ratio of the lens galaxy for ID81 and ID130 (Table$\,$2) \cite{SOM} to those measured in the Sloan Lens Advanced Camera for Surveys (ACS) (fig.$\,$4) \cite{Auger09}, which cover a similar redshift range (z$\sim$0.1 to 0.3). The agreement with the average trend revealed by SLACS confirms that our lens selection method is not biased to lensing ellipticals with atypical luminosities. Moreover, the location of ID130 in Fig.$\,$4 indicates that our 
selection method can probe lower masses and lower luminosity 
lens galaxies than those sampled by SLACS, thus offering a wider range in lens properties to be investigated. \\
The best fit SED to the submillimeter/millimeter photometry for each of the
five background sources give infrared luminosities L$_{\rm
  IR}\ge\sim$3$\times$10$^{13}\,$L$_{\odot}$ (Table$\,$2), which would
classify these objects as Hyper Luminous Infra-Red galaxies (HLIRGs;
L$_{\rm IR}\ge$10$^{13}\,$L$_{\odot}$). However, a correction for magnification
because of lensing will reduce these values by a factor of 10 or
greater. For example, assuming that the light distribution of the
background source is described by a Gaussian profile with a full width
at half maximum (FWHM) of 0.2$\,$arcseconds [which is consistent with the
physical extension of the background galaxy in \cite{Swinbank10}], the
best-fit lens model (fig.$\,$S9) \cite{SOM} predicts a total amplifications of $\sim$19 and $\sim$6
for ID81 and ID130, respectively. Typical amplifications of 8 to 10 are also suggested by \cite{Negrello07}, therefore, it is more likely that these sources are ULIRGs. \\
These results already provide constraints for models of the
formation and evolution of massive galaxies at high redshift. The fact that many (if not all) of the brightest SMGs detected in the H-ATLAS
SDP field are
amplified by lensing, implies that un-lensed $z>1$ star-forming
galaxies with flux densities more than 100$\,$mJy at 500$\,\mu$m are
rare, with $\leq4.6$ of them per 14.4$\,$deg$^{-2}$, at 99$\%$
probability (assuming Poisson statistics). This translates into a
0.32$\,$deg$^{-2}$ upper limit on the surface density of these
sources.  The same limit should translate to the abundance of HLIRGs 
with L$_{\rm IR}>$5$\times$10$^{13}\,$L$_{\odot}$at $z < 4$, because they
would also have 500-$\mu$m flux densities above $100\,$mJy, which has possible implications 
for the role of feedback during the formation of the most massive galaxies in the universe. 
By extrapolating our SDP findings to the full H-ATLAS field, 
we predict a total sample of more than 100 bright lensed sources, with which we can further improve this constraint.

\phantom{\cite{acknowledgments}} \bibliography{1193420Revisedtext}

\begin{thebibliography}{10}

\bibitem{notes1}
{When multiple images of the same background source are formed, the event is
  known as strong gravitational lensing} .

\bibitem{Marshall07}
P.~J. {Marshall}, {\it et~al.\/}, {\it Astrophys. J.\/} {\bf 671}, 1196 (2007).

\bibitem{Stark08}
D.~P. {Stark}, {\it et~al.\/}, {\it Nature\/} {\bf 455}, 775 (2008).

\bibitem{Swinbank10}
A.~M. {Swinbank}, {\it et~al.\/}, {\it Nature\/} {\bf 464}, 733 (2010).

\bibitem{Smail97}
I.~{Smail}, R.~J. {Ivison}, A.~W. {Blain}, {\it Astrophys. J. L.\/} {\bf 490},
  L5+ (1997).

\bibitem{Blain99}
A.~W. {Blain}, J.~{Kneib}, R.~J. {Ivison}, I.~{Smail}, {\it Astrophys. J. L.\/}
  {\bf 512}, L87 (1999).

\bibitem{RR91}
M.~{Rowan-Robinson}, {\it et~al.\/}, {\it Nature\/} {\bf 351}, 719 (1991).

\bibitem{Smail02}
I.~{Smail}, R.~J. {Ivison}, A.~W. {Blain}, J.~{Kneib}, {\it Mon. Not. R.
  Astron. Soc.\/} {\bf 331}, 495 (2002).

\bibitem{Vieira10}
J.~D. {Vieira}, {\it et~al.\/}, {\it Astrophys. J.\/} {\bf 719}, 763 (2010).

\bibitem{Pilbratt10}
G.~L. {Pilbratt}, {\it et~al.\/}, {\it Astron. \& Astrophys.\/} {\bf 518}, L1+
  (2010).

\bibitem{Blain96}
A.~W. {Blain}, {\it Mon. Not. R. Astron. Soc.\/} {\bf 283}, 1340 (1996).

\bibitem{Perrotta02}
F.~{Perrotta}, C.~{Baccigalupi}, M.~{Bartelmann}, G.~{De Zotti}, G.~L.
  {Granato}, {\it Mon. Not. R. Astron. Soc.\/} {\bf 329}, 445 (2002).

\bibitem{Perrotta03}
F.~{Perrotta}, {\it et~al.\/}, {\it Mon. Not. R. Astron. Soc.\/} {\bf 338}, 623
  (2003).

\bibitem{Negrello07}
M.~{Negrello}, {\it et~al.\/}, {\it Mon. Not. R. Astron. Soc.\/} {\bf 377},
  1557 (2007).

\bibitem{Coppin06}
K.~{Coppin}, {\it et~al.\/}, {\it Mon. Not. R. Astron. Soc.\/} {\bf 372}, 1621
  (2006).

\bibitem{Chapman05}
S.~C. {Chapman}, A.~W. {Blain}, I.~{Smail}, R.~J. {Ivison}, {\it Astrophys.
  J.\/} {\bf 622}, 772 (2005).

\bibitem{Serjeant05}
S.~{Serjeant}, D.~{Harrison}, {\it Mon. Not. R. Astron. Soc.\/} {\bf 356}, 192
  (2005).

\bibitem{DeZotti05}
G.~{de Zotti}, {\it et~al.\/}, {\it Astron. \& Astrophys.\/} {\bf 431}, 893
  (2005).

\bibitem{Koopmans06}
L.~V.~E. {Koopmans}, T.~{Treu}, A.~S. {Bolton}, S.~{Burles}, L.~A. {Moustakas},
  {\it Astrophys. J.\/} {\bf 649}, 599 (2006).

\bibitem{Browne01}
I.~W.~A. {Browne}, {The Class Collaboration}, {\it Gravitational Lensing:
  Recent Progress and Future Go\/}, {T.~G.~Brainerd \& C.~S.~Kochanek}, ed.
  (2001), vol. 237 of {\it Astronomical Society of the Pacific Conference
  Series\/}, pp. 15--+.

\bibitem{Oguri06}
M.~{Oguri}, {\it et~al.\/}, {\it Astron. J.\/} {\bf 132}, 999 (2006).

\bibitem{Faure08}
C.~{Faure}, {\it et~al.\/}, {\it Astrophys. J. S.\/} {\bf 176}, 19 (2008).

\bibitem{Weiss09}
A.~{Wei{\ss}}, {\it et~al.\/}, {\it Astrophys. J.\/} {\bf 707}, 1201 (2009).

\bibitem{Devlin09}
M.~J. {Devlin}, {\it et~al.\/}, {\it Nature\/} {\bf 458}, 737 (2009).

\bibitem{Eales10}
S.~{Eales}, {\it et~al.\/}, {\it Publ. Astron. Soc. Pac.\/} {\bf 122}, 499
  (2010).

\bibitem{Griffin10}
M.~J. {Griffin}, {\it et~al.\/}, {\it Astron. \& Astrophys.\/} {\bf 518}, L3+
  (2010).

\bibitem{Pascale10}
E.~{Pascale}, {\it submitted; preprint available at
  http://xxx.lanl.gov/abs/astro-ph/1010.5782\/} .

\bibitem{Poglitsch10}
A.~{Poglitsch}, {\it et~al.\/}, {\it Astron. \& Astrophys.\/} {\bf 518}, L2+
  (2010).

\bibitem{Ibar10}
E.~{Ibar}, {\it submitted; preprint available at
  http://xxx.lanl.gov/abs/astro-ph/1009.0262\/} .

\bibitem{Driver10}
S.~{Driver}, {\it submitted; preprint available at
  http://xxx.lanl.gov/abs/astro-ph/1009.0614\/} .

\bibitem{Rigby10}
E.~E. {Rigby}, {\it submitted; preprint available at
  http://xxx.lanl.gov/abs/astro-ph/1010.5787\/} .

\bibitem{Clements10}
D.~L. {Clements}, {\it et~al.\/}, {\it Astron. \& Astrophys.\/} {\bf 518}, L8+
  (2010).

\bibitem{Baes10}
M.~{Baes}, {\it et~al.\/}, {\it Astron. \& Astrophys.\/} {\bf 518}, L39+
  (2010).

\bibitem{Gonzalez-Nuevo10}
J.~{Gonz{\'a}lez-Nuevo}, {\it et~al.\/}, {\it Astron. \& Astrophys.\/} {\bf
  518}, L38+ (2010).

\bibitem{SOM}
{Supporting online material (SOM)} .

\bibitem{Smith10}
D.~J.~B. {Smith}, {\it et~al.\/}, {\it submitted; preprint available at
  http://xxx.lanl.gov/abs/astro-ph/1007.5260\/} .

\bibitem{Hughes02}
D.~H. {Hughes}, {\it et~al.\/}, {\it Mon. Not. R. Astron. Soc.\/} {\bf 335},
  871 (2002).

\bibitem{Aretxaga03}
I.~{Aretxaga}, {\it et~al.\/}, {\it Mon. Not. R. Astron. Soc.\/} {\bf 342}, 759
  (2003).

\bibitem{Weib09}
A.~{Wei{\ss}}, {\it et~al.\/}, {\it Astrophys. J. L.\/} {\bf 705}, L45 (2009).

\bibitem{Daddi09}
E.~{Daddi}, {\it et~al.\/}, {\it Astrophys. J. L.\/} {\bf 695}, L176 (2009).

\bibitem{Naylor03}
B.~J. {Naylor}, {\it et~al.\/}, {\it Society of Photo-Optical Instrumentation
  Engineers (SPIE) Conference Series\/}, {T.~G.~Phillips \& J.~Zmuidzinas}, ed.
  (2003), vol. 4855 of {\it Presented at the Society of Photo-Optical
  Instrumentation Engineers (SPIE) Conference\/}, pp. 239--248.

\bibitem{Bradford09}
C.~M. {Bradford}, {\it et~al.\/}, {\it Astrophys. J.\/} {\bf 705}, 112 (2009).

\bibitem{Lupu10}
R.~{Lupu}, {\it submitted; preprint available at
  http://xxx.lanl.gov/abs/astro-ph/1009.5983\/} .

\bibitem{Harris07}
A.~I. {Harris}, {\it et~al.\/}, {\it From Z-Machines to ALMA: (Sub)Millimeter
  Spectroscopy of Galaxies\/}, {A.~J.~Baker, J.~Glenn, A.~I.~Harris,
  J.~G.~Mangum, \& M.~S.~Yun }, ed. (2007), vol. 375 of {\it Astronomical
  Society of the Pacific Conference Series\/}, pp. 82--+.

\bibitem{notes2}
{The Zpectrometer frequency coverage allows it to search for CO J=1$-$0
  transition over the redshift range 2.2 to 3.5 simultaneously} .

\bibitem{Frayer10}
D.~{Frayer}, {\it submitted; preprint available at
  http://xxx.lanl.gov/abs/astro-ph/1009.2194\/} .

\bibitem{notes5}
{Only g and i-band photometry can be obtained for each of the two galaxies
  detected in ID17. For this reason, we did not attempt any fit to the
  optical/NIR SED of this H-ATLAS source} .

\bibitem{DaCunha08}
E.~{da Cunha}, S.~{Charlot}, D.~{Elbaz}, {\it Mon. Not. R. Astron. Soc.\/} {\bf
  388}, 1595 (2008).

\bibitem{daCunha10}
E.~{da Cunha}, {\it et~al.\/}, {\it in press; preprint available at
  http://xxx.lanl.gov/abs/astro-ph/1008.2000\/} .

\bibitem{notes4}
{The visual extinction parameter, $A_{V}$, is the difference between the
  observed V-band magnitude and the intrinsic (dust-free) V-band magnitude of
  the galaxy} .

\bibitem{Faber76}
S.~M. {Faber}, R.~E. {Jackson}, {\it Astrophys. J.\/} {\bf 204}, 668 (1976).

\bibitem{Willott03}
C.~J. {Willott}, S.~{Rawlings}, M.~J. {Jarvis}, K.~M. {Blundell}, {\it Mon.
  Not. R. Astron. Soc.\/} {\bf 339}, 173 (2003).

\bibitem{Huang98}
J.~{Huang}, L.~L. {Cowie}, G.~A. {Luppino}, {\it Astrophys. J.\/} {\bf 496}, 31
  (1998).

\bibitem{Gudehus91}
D.~H. {Gudehus}, {\it Astrophys. J.\/} {\bf 382}, 1 (1991).

\bibitem{Auger09}
M.~W. {Auger}, {\it et~al.\/}, {\it Astrophys. J.\/} {\bf 705}, 1099 (2009).

\bibitem{acknowledgments}
{Herschel is an ESA space observatory with science instruments provided by
  European-led Principal Investigator consortia and with important
  participation from NASA. US participants in H-ATLAS acknowledge support from
  NASA through a contract from JPL. This work was supported by the Science and
  Technology Facilities Council (grants PP/D002400/1 and ST/G002533/1) and
  studentship SF/F005288/1. We thank Agenzia Spaziale Italiana for funding
  through contract No. I/016/07/0 COFIS and ASI/Istituto Nazionale di
  Astrofisica agreement I/072/09/0 for the Planck Low Frequency Instrument
  (LFI) Activity of Phase E2. Research supported in part by Consejo Nacional de
  Ciencia y Tecnologia (CONACyT) grants 39953-F and 39548-F. The W.M. Keck
  Observatory is operated as a scientific partnership among the California
  Institute of Technology, the University of California and NASA. The
  Observatory was made possible by the generous financial support of the W.M.
  Keck Foundation. The Submillimeter Array is a joint project between the
  Smithsonian Astrophysical Observatory and the Academia Sinica Institute of
  Astronomy and Astrophysics and is funded by the Smithsonian Institution and
  the Academia Sinica. IRAM is supported by Institut National des Sciences de
  l'Univers (INSU)/CNRS (France), Max Planck Society (MPG) (Germany) and
  Instituto Geografico Nacional (Spain). Z-spec was supported by NSF grant
  AST-0807990 to J. A. and by the CSO NSF Cooperative Agreement AST-0838261.
  Support was provided to J. K. by an NSF Graduate Research Fellowship. Z-spec
  was constructed under NASA SARA grants NAGS-11911 and NAGS-12788 and an NSF
  Career grant (AST-0239270) and a Research Corporation Award (RI0928) to J.
  G., in collaboration with the JPL, California Institute of Technology, under
  a contract with NASA. Construction of and observations with the Zpectrometer
  have been supported by NSF grants AST-0503946 and AST-0708653. The NRAO is a
  facility of the NSF operated under cooperative agreement by Associated
  Universities. The optical spectroscopic redshift of ID130 was derived from
  observations obtained with the Apache Point Observatory 3.5-meter telescope,
  which is owned and operated by the Astrophysical Research Consortium. The
  optical spectroscopic redshifts of ID9 and ID11 were obtained with the
  William Herschel Telescope which is operated on the island of La Palma by the
  Isaac Newton Group in the Spanish Observatorio del Roque de los Muchachos of
  the Instituto de Astrofísica de Canarias. For the use of Keck, SMA and CSO,
  the authors wish to recognize and acknowledge the very important cultural
  role and reverence that the summit of Mauna Kea has always had within the
  indigenous Hawaiian community. We are most fortunate to have the opportunity
  to conduct observations from this mountain} .

\end{thebibliography}
\bibliographystyle{Science}

\clearpage

$~$ \\
{\LARGE {\bf Tables}} \\

\begin{table*}[!bht]
  \vspace{0.0cm}
  \begin{center}\caption{Photometric and spectroscopic redshifts of the five
lens candidates. Spectroscopic redshifts were derived from optical
lines for the lens [$z_{\rm spec}^{\rm (opt)}$] and from CO lines for
the background source [$z_{\rm spec}^{\rm (CO)}$]. Photometric
redshifts are based on UV/optical/NIR photometry for the lens [$z_{\rm
  ot}^{\rm (opt)}$] and H-ATLAS plus SMA and Max-Planck Millimeter Bolometer
(MAMBO) photometry for the background source[$z_{\rm phot}^{\rm
  (submillimeter/millimeter)}$; using the photometric redshift code of
\cite{Hughes02,Aretxaga03}]. The quoted errors on the redshifts
correspond to a 68$\%$ confidence interval.
}
    \vspace{+0.5cm}
    \footnotesize
    \begin{tabular}{ccccc}
      \hline 
      \multicolumn{1}{l}{SDP ID} 
      & \multicolumn{1}{c}{$z_{\rm phot}^{\rm (opt)}$} 
      & \multicolumn{1}{c}{$z_{\rm spec}^{\rm (opt)}$} 
      & \multicolumn{1}{c}{$z_{\rm phot}^{\rm (sub-mm/millimeter)}$} 
      & \multicolumn{1}{c}{$z_{\rm spec}^{\rm (CO)}$} 
      \\
      \hline
      \hline
      9    &  0.679$\pm$0.057       &  $-$                  & 1.4$_{-0.4}^{+0.3}$&  1.577$\pm$0.008$^{(1)}$  \\  
      11   &  0.72$\pm$0.16  &  0.7932$\pm$0.0012$^{(2)}$     & 1.9$_{-0.3} ^{+0.4}$ & 1.786$\pm$0.005$^{(1)}$  \\
      17   &  0.77$\pm$0.13         &  0.9435$\pm$0.0009$^{(2)}$  & 2.0$_{-0.3}^{+0.4}$ &  0.942$\pm$0.004 \& 2.308$\pm$0.011$^{(1)}$  \\ 
      81   &  0.334$\pm$0.016       &  0.2999$\pm$0.0002$^{(3)}$  & 2.9$_{-0.3}^{+0.2}$ &  3.037$\pm$0.010$^{(1)}$  \\
      &                         &                     &                          & 3.042$\pm$0.001$^{(4),(5)}$ \\
      130  &  0.239$\pm$0.021       &  0.2201$\pm$0.002$^{(6)}$                  & 2.6$_{-0.2}^{+0.4}$ &  2.625$\pm$0.001$^{(4)}$   \\
           &                       &                                           &       &     2.6260$\pm$0.0003$^{(5)}$ \\
      \hline
      \multicolumn{5}{l}{$^{(1)}$ Datum is from CSO/Z-Spec \cite{Lupu10}} \\
      \multicolumn{5}{l}{$^{(2)}$ Datum is from the William Herschel Telescope \cite{SOM}} \\
      \multicolumn{5}{l}{$^{(3)}$ Datum is from SDSS} \\
      \multicolumn{5}{l}{$^{(4)}$ Datum is from GBT/Zpectrometer \cite{Frayer10}} \\
      \multicolumn{5}{l}{$^{(5)}$ Datum is from PdBI \cite{SOM}} \\
      \multicolumn{5}{l}{$^{(6)}$ Datum is from the Apache Point Observatory \cite{SOM}} \\
    \end{tabular} 
  \end{center}
\end{table*}

\begin{table*}[!bht]
  \vspace{0.0cm}
  \begin{center}\caption{Derived parameters for the five lens
candidates. Estimated mass in stars (M$_{\star}$) and Star Formation
Rate (SFR) of the foreground galaxy derived from the best-fit to the
UV/optical/near-IR part of the SED; the Einstein radius measured from
the SMA images ($\theta_{\rm E}$); mass within the Einstein radius (M$_{\rm E}$) estimated from
$\theta_{\rm E}$; line-of-sight stellar velocity dispersion
($\sigma^{\rm FJ}_{\rm v}$) derived from the Faber-Jackson relation
and the B-band luminosity produced by the best-fit to the
UV/optical/NIR SED; Einstein radius ($\theta^{\rm FJ}_{\rm E}$)
calculated from $\sigma^{\rm FJ}_{\rm v}$; infrared luminosity of the
background source (L$_{\rm IR}$), without correction for
magnification, derived by fitting the submillimeter/millimeter part of the
SED and the upper limits at optical and NIR wavelengths
(Fig. 2); and visual extinction parameter (A$_{V}$) inferred for the
background galaxy. All the quoted errors correspond to a 68 per cent
confidence interval. For ID17 only the infrared luminosity and the
extinction parameter of the background source are quoted because the
lensing mass probably consists of two galaxies that can only be
disentangled in the Keck images. The symbols M$_{\odot}$ and $L_{\odot}$ denote the total mass and the total luminosity of the Sun, respectively, and correspond to $M_{\odot}=1.99\times10^{30}\,$kg and $L_{\odot}=3.839\times10^{33}\,$erg$\,$s$^{-1}$. Dashes indicate lack of constraints.}
    \begin{minipage}{15cm}
    \vspace{+0.5cm}
    \scriptsize
    \begin{tabular}{c|cccccc|cc}
      \hline 
      \multicolumn{1}{l}{H-ATLAS} 
      & \multicolumn{1}{|c}{log(M$_{\star}$)} 
      & \multicolumn{1}{c}{log(SFR)} 
      & \multicolumn{1}{c}{$\theta_{\rm E}$} 
      & \multicolumn{1}{c}{log(M$_{\rm E}$)} 
      & \multicolumn{1}{c}{$\sigma^{\rm FJ}_{\rm v}$}
      & \multicolumn{1}{c}{$\theta^{\rm FJ}_{\rm E}$} 
      & \multicolumn{1}{|c}{log(L$_{\rm IR}$)} 
      & \multicolumn{1}{c}{A$_{V}$} \\
      ID & (M$_{\odot}$) & (M$_{\odot}\,$yr$^{-1}$)  &  (arcsec) & (M$_{\odot}$) & (km$\,$sec$^{-1}$) & (arcsec) & (L$_{\odot}$$^{a}$) & \\
      \hline
        \hline
     9    & 10.79$_{-0.11}^{+0.16}$ & -0.51$_{-0.27}^{+0.20}$  &  $-$  &  $-$  &  232$_{-56}^{+75}$ & 0.77$_{-0.34}^{+0.49}$ & 13.48$_{-0.06}^{+0.07}$ & 6.7$_{-1.0}^{+1.5}$ \\
     11   & 11.15$_{-0.10}^{+0.09}$ & -0.08$_{-0.24}^{+0.18}$  & $-$ & $-$ & 258$_{-62}^{+82}$ & 0.91$_{-0.40}^{+0.59}$ & 13.61$_{-0.06}^{+0.06}$ & 5.1$_{-0.7}^{+1.6}$ \\
     17   &          $-$          &    $-$                 & $-$ &  $-$ & $-$ & $-$  & 13.57$_{-0.06}^{+0.08}$ & 5.3$_{-0.6}^{+1.4}$ \\
     81   & 11.17$_{-0.08}^{+0.04}$ & -1.66$_{-0.14}^{+0.46}$   & 1.62$\pm0.02$ & 11.56$\pm0.01$ & 242$_{-58}^{+77}$ & 1.51$_{-0.67}^{+0.98}$ & 13.71$_{-0.07}^{+0.07}$ & 3.5$_{-0.3}^{+3.4}$ \\
     130  & 10.65$_{-0.08}^{+0.06}$ & -1.17$_{-0.58}^{+0.39}$   &  0.59$\pm$0.02 & 10.57$\pm0.04$ & 174$_{-42}^{+55}$ & 0.81$_{-0.36}^{+0.52}$  & 13.45$_{-0.09}^{+0.08}$ &  1.9$_{-0.3}^{+0.3}$ \\
     \hline
    \end{tabular} 
    \end{minipage}
  \end{center}
\end{table*}

\clearpage

$~$ \\
{\LARGE {\bf Figures}} \\

\setcounter{figure}{0}\renewcommand{\tablename}{Figure}
\begin{figure}[!bht]
  \hspace{-1.0cm}\includegraphics[height=12.0cm,width=14.0cm]{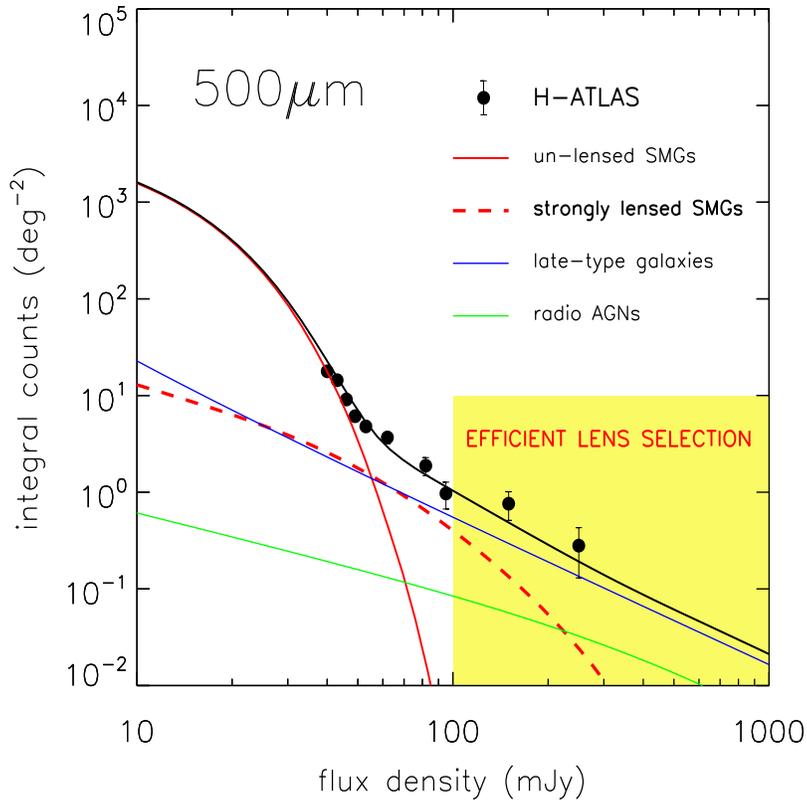} \\
  \vskip-1.5truecm
  \caption{{\normalsize Selection of gravitational lenses at submillimeter
wavelengths. The 500$\,\mu$m source counts consist of three different
populations \cite{Negrello07}: high-redshift SMGs; 
lower redshift late type (starburst plus normal spiral) galaxies; and radio
sources powered by active galactic nuclei. Strongly lensed SMGs
dominate over unlensed SMGs at very bright fluxes where the count of
un-lensed SMGs falls off dramatically (yellow shaded region). The data
points are from H-ATLAS \cite{Clements10}.}} 
\end{figure}

\setcounter{figure}{1}\renewcommand{\tablename}{Figure}
\begin{figure*} 
  \hspace{-0.0cm}\includegraphics[height=14.0cm,width=16.5cm]{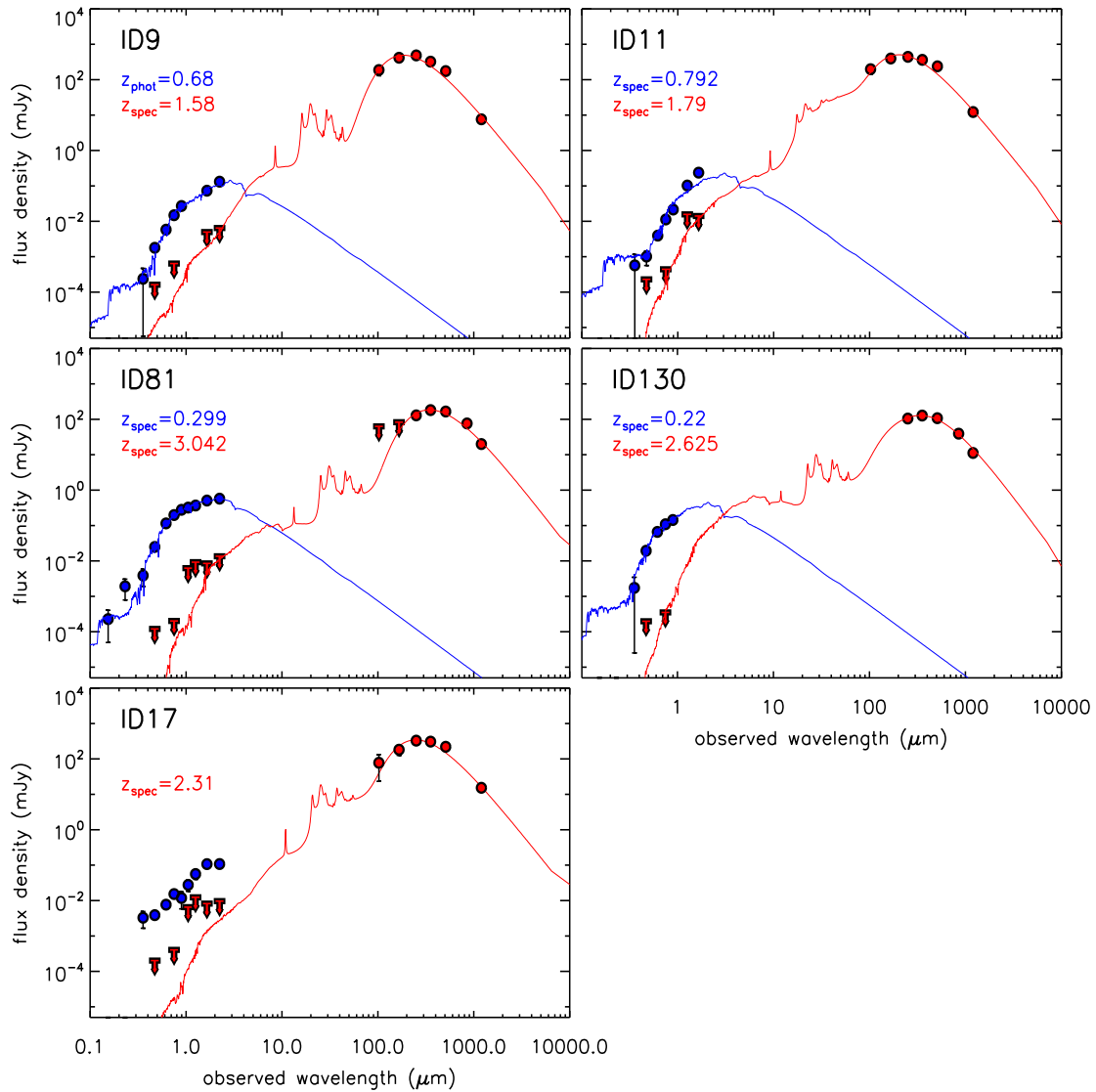} \\
  \vskip+0.5truecm
  \caption{{\normalsize Spectra of the gravitational lens candidates.  The UV,
optical and NIR data points (blue dots) are from {\it GALEX}, SDSS and
UKIDSS LAS, respectively.  The submillimeter/millimeter data points (red
dots) are from PACS/Herschel, SPIRE/Herschel, SMA and
Max-Planck Millimeter Bolometer (MAMBO)/IRAM. Upper limits at PACS/Herschel wavelengths are shown at
3$\,\sigma$. ID130 lies outside the region covered by PACS. The
photometric data were fitted using SED models from
\cite{DaCunha08}. The background source, responsible for the submillimeter
emission, is a heavily dust obscured star-forming galaxy (red solid
curve), whereas the lens galaxy, which is responsible for the UV/optical and NIR
part of the spectrum, is caracterized by passive stellar evolution.}} 
\end{figure*}

\setcounter{figure}{2}\renewcommand{\tablename}{Figure}
\begin{figure*} 
  \hspace{-0.0cm}\includegraphics[height=14.0cm,width=16.5cm]{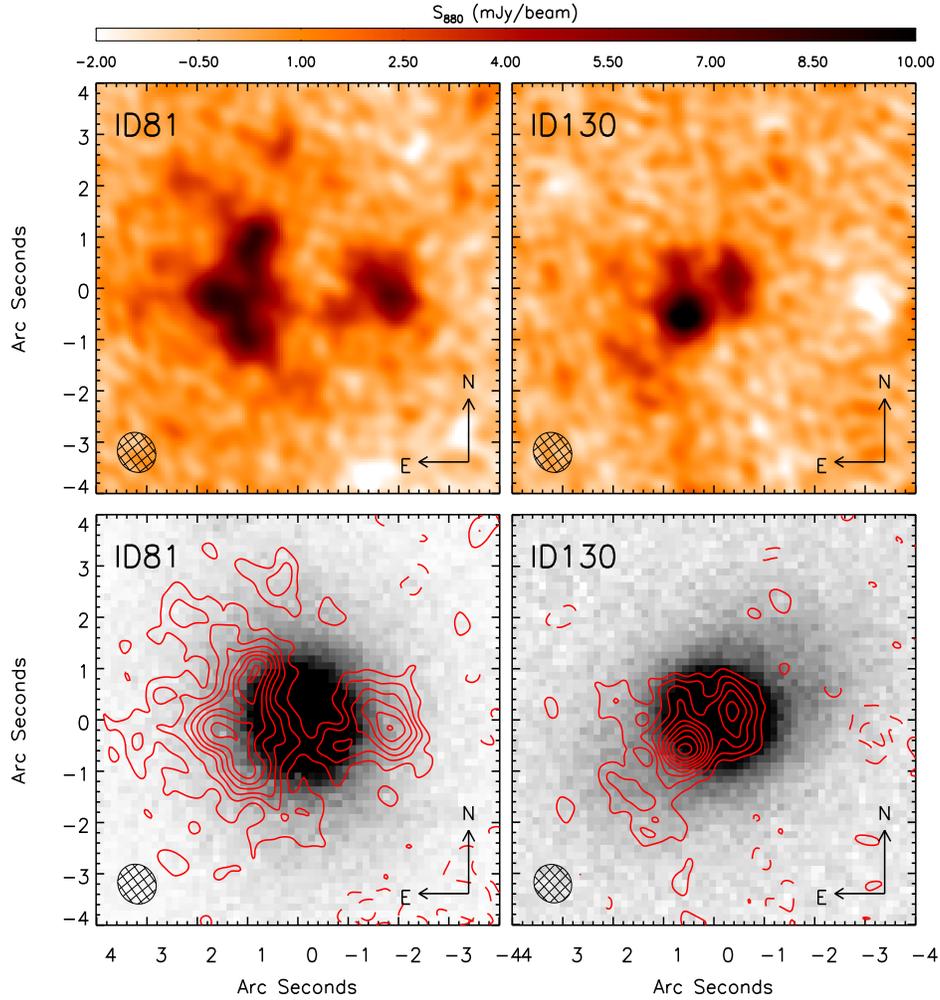} \\
  \vskip+0.5truecm
  \caption{{\normalsize Submillimeter and optical follow-up imaging of ID81 and
ID130. The SMA images of ID81 and ID130 are shown in the top panels,
centered on the optical counterpart, and were obtained by combining
the visibility data from very-extended, compact and sub-compact
configuration observations. The Keck i-band image of ID81 and ID130
are shown in the bottom panels with the SMA
contours superimposed (in red). The contours are in steps of -2$\sigma$, 2$\sigma$,
4$\sigma$, 6$\sigma$, 8$\sigma$, 10$\sigma$... , with
$\sigma=0.6\,$mJy/beam. The SMA synthesized beam is shown in the
bottom-left corner.}} 
\end{figure*}

\setcounter{figure}{3}\renewcommand{\tablename}{Figure}
\begin{figure*} 
  \hspace{-0.0cm}\includegraphics[height=9.5cm,width=12.0cm]{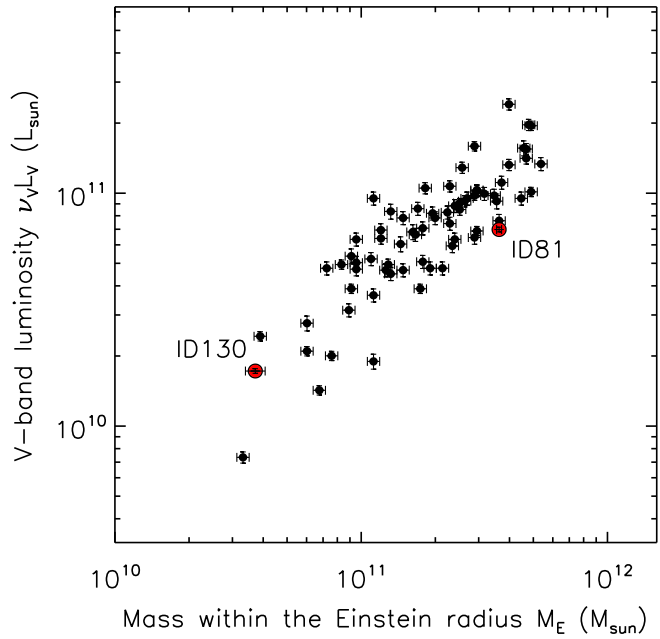} \\
  \vskip+0.5truecm
  \caption{{\normalsize Relationship between mass and luminosity for the
lensing galaxy in ID81 and ID130. The rest-frame $V-$band luminosity
was derived from the best-fit SED to the UV/optical/NIR photometric
data; the mass within the Einstein radius is that measured directly
from the SMA images. The light versus mass relation inferred for ID81
and ID130 (red dots) is consistent with that observed for the SLACS
lenses [black dots, from \cite{Auger09} assuming an uncertainty of
0.025dex in their mass estimates].}} 
\end{figure*}


\clearpage

$~$ \\
{\bf {\LARGE Supporting Online Material}} \\

$~$ \\
{\bf MAMBO observations}
$~$ \\
\\
The five H-ATLAS/SDP lens candidates, i.e. ID9, ID11, ID17, ID81 and ID130,
were observed on 2010 March 9 and 10, with MAMBO, at the IRAM 30 meter telescope on Pico Veleta, in Diretcor's Discretionary Time (DDT). 
The MAMBO array consists of 117
bolometer elements and operates at a central frequency of 250$\,$GHz,
corresponding to 1.2$\,$mm. The beam size (FWHM$\,\approx\,$11 arcsec) of
MAMBO ensures that the true dust emission at 1.2$\,$mm is
obtained if the source is not more extended
than a few arcseconds.
Each science target was observed in the photometric mode
(``on-off'') of MAMBO which is based on the chop-nod technique and
placing the target on a reference bolometer element (on-target
channel). The total observing time was 1.5 hours. Data were reduced
using MOPSIC, and the current version of MOPSI (Zylka 1998, The MOPSI
Cookbook (Bonn: MPIfR)).

$~$ \\
$~$ \\
{\bf Keck observations}
$~$ \\
\\
The imaging observations were acquired on 10 March 2010 using the
dual-arm Low Resolution Imaging Spectrometer [LRIS; ({\it 1, 2})] on the 10-m Keck I telescope.  Each
target field received simultaneous $110\times3\,$s integrations with
the $g$-filter and $60\times3$s integrations with the $i$-filter using
the blue and red arms of LRIS. A $\sim20^{\prime\prime}$ dither
pattern was employed to generate on-sky flat field frames when
incorporating all five fields.  In addition, $1\,$s integrations were
acquired in the $g$- and $i$-filters for photometric calibration of
bright stars in each field. The seeing FWHM for the science exposures
was $\sim0.8-0.9^{\prime\prime}$. The data were reduced using IDL
routines and combined and analysed using standard IRAF tasks.

$~$ \\
$~$ \\
{\bf SMA observations}
$~$ \\
\\
Observations of two H-ATLAS/SDP sources, ID81 and ID130, were obtained at
880$\,\mu$m using the only current high resolution
submillimeter facility in the world, the SMA. 
The SMA is an interferometer located near the summit of Mauna Kea,
Hawaiim and consists of eight 6$\,$m diameter radio telescopes.  The two
H-ATLAS sources were observed in Director's Discretionary Time (from February to May 2010) in three separate configurations, with baselines spanning
a spatial range from 6 to 509 meters, over a total of
4 observing periods (Table S3). \\
Target observations from each period were interspersed with
observations of calibration sources, quasars J0909+013 and J0825+031
(phase) and J0730-116 and J0854+201 (amplitude).  The phase
calibration targets were typically observed every 7 to 15 minutes,
depending upon
configuration (a faster cycle was used for the larger configurations). \\
Calibration of the complex visibility data was performed within the
SMA's MIR package, a suite of IDL-based routines designed for use with
SMA data. The initial opacity correction was obtained through application
of system temperature to the raw visibility data, a standard practice.
Further complex gain corrections, to remove both atmospheric and
instrumental amplitude and phase variations, were measured using the
calibration quasars, which appear as point sources to the
interferometer.  Calibration of the absolute flux density scale was
performed using measurements of Titan, whose continuum and line
structure is known to
within about 5\% at submillimeter/millimeter wavelengths. \\
The resulting calibrated visibility data for each source were combined
and imaged within the NRAO Astronomical Image Processing System
(AIPS). Photometry obtained for the SMA images along with those from PACS, SPIRE, MAMBO are given in table S1.

$~$ \\
$~$ \\
{\bf Plateau de Bure Interferometer observations}
$~$ \\
\\
The H-ATLAS/SDP sources ID81 and ID130 were observed in the CO J=3-2 and CO
J=5-4 lines with the IRAM Plateau de Bure Interferometer ({\it 3}). Both sources were observed in excellent atmospheric conditions and
with the full sensitivity of the six-element array. The observing
frequencies were based on the redshifts provided by the CSO/Z-spec
spectrometer. The receiver bandwidth was adjusted for maximum sensitivity
and the observing frequencies centered in the 1 GHz baseband of the
narrowband correlator. Observations of ID81 were made on March 22, 2010 for
an effective integration time of 22 min and 14 min, respectively, for the CO
J=3-2 and J=5-4 lines. The RF calibration was measured on 3C84, and
amplitude and phase calibrations were made on 0823+033. The J=3-2 and J=5-4
transitions in ID130 were observed on March 26 and April 16, 2010,
respectively, for an effective integration time of 74 min and 32 min. The RF
calibration was measured on 3C273, and amplitude and phase calibrations were
made on 0906+015. The absolute flux calibration scales for ID81 and ID130
were established using as primary calibrator MWC349. Data reduction and
calibration were made using the GILDAS software package in the standard
antenna based mode.

$~$ \\
$~$ \\
{\bf Optical spectroscopic observations}
$~$ \\
\\
Optical spectroscopic observations of ID11 and ID17 were made using the ISIS
double-arm spectrograph on the 4.2-m William Herschel Telescope (WHT). The
R158B and R158R gratings were used to provide wavelength coverage
across the entire optical spectrum, split by a dichroic at
$\sim$5300\,\AA. Four 900-second exposures were taken of each source
in a standard `ABBA' pattern, nodding the telescope along the slit by
10 arcseconds between the first and second exposures, and back to the
original position between the third and fourth integrations. This
allowed initial sky subtraction to be performed by simply subtracting
the 'A' frames from the 'B' frames. Additional sky subtraction was
performed by subtracting the median value of each row, and then the
positive and negative beams were aligned and coadded. Wavelength
calibration was performed using observations of arc lamps taken with
the same set-up. A one-dimensional spectrum was then optimally
extracted. The spectra were taken through thin cloud and therefore no attempt
has been made to flux-calibrate them. There was very little signal in
the blue arms and so only the red-arm spectra are presented here. The
redshifts of the two sources were determined by cross-correlation with
template spectra. All reduction steps were undertaken using the IRAF
package. The resulting spectra are shown in Fig. S3. The spectrum of ID11 reveals absorption lines associated to singly ionised calcium Ca H+K (rest-frame wavelengths: 3968.5$\,\AA$ for H-line and 3933.7$\,\AA$ for K-line) and the 4000$\,\AA$ break feature (rest-frame wavelength 4000$\,\AA$) at $z=0.793$, while the spectrum of ID17 shows the emission from oxygen doublet [OII]3727 (rest-frame wavelengths 3726-3729$\,\AA$) and the 4000$\,\AA$ break feature at $z=0.944$. In both spectra the absorption feature observed at $\sim7600\,\AA$ is due to the Earth's atmosphere. \\
A 30-minute exposure of ID130 was taken on May 15, 2010, with the Apache Point Observatory's 3.5-meter telescope and the DIS [Dual Imaging Spectrograph, ({\it 4})] long-slit spectrograph through medium clouds at an average airmass of 1.5.  The spectrum was processed by subtracting the detector bias, dividing by a flat-field frame to correct for variable pixel response, performing distortion correction to align the spectrum in the wavelength and spatial directions, subtracting the sky flux determined from parts of the slit containing no sources, and applying a wavelength calibration by reference to emission lines from a Helium-Neon-Argon calibration lamp.  Two emission lines in the spectrum (Fig. S4) were identified as [O II]3727 and [Ne III]3869 (rest-frame wavelengths 3869$\,\AA$) from the ratio of their observed wavelengths. From the ratio of their observed to emitted wavelengths the redshift of the galaxy was determined to be $z=$0.2201$\pm$0.002.

\clearpage

$~$ \\
{\bf Modelling with GALFIT}
$~$ \\
\\
GALFIT ({\it 5}) is a publicly available two-dimensional non-linear fitting algorithm, which allows galaxy images to be modelled with one or multiple analytical light profiles. Each profile is constrained by a function and a set of parameters. GALFIT convolves the profiles with a user supplied point spread function, in this case empirical point spread functions constructed using nearby stars, and then performs a least-squares minimisation. No hard or soft constraints were applied to the fitting parameters to avoid any prior on the galaxy morphological type. For ID9 and ID11 single Sersic profiles resulted in a reduced $\chi^2$ close to 1.0 (see table S4 for the best fit parameters). ID17 was fitted with two Sersic component, assuming two lensing galaxies. The resulting Sersic indices were both less than 1 (see table S4). For ID81 and ID130 two components were necessary to achieve a satisfactory fit, with a clean residual. The best fits were obtained using a combination of a compact elliptical Sersic core plus an exponential disk. No detectable background structure was revealed after subtracting the models, which shows the background galaxy is below the optical detection limit. Postage stamp images of ID9, ID11, ID17, ID81 and ID130 are shown in Figs$\,$S5 and S6, together with the corresponding best-fit models and residuals, while Fig.$\,$S7 shows the individual GALFIT components for ID81 and ID130.\\
To derive photometric upper limits, we performed random
aperture photometry on the i- and g-band Keck maps, using a 1.5
arcsecond radius. This radius was chosen to correspond with the structure visible in the SMA images for ID81 and ID130, which extends to regions with radii of approximately 1$-$1.5 arcseconds. The resulting flux distributions were fitted with
Gaussians and the 3$-\sigma$ upper limits are presented in Table S4.

\clearpage
$~$ \\
{\bf Mass estimate from lensing}
$~$ \\
\\
The Einstein radius of a strong galaxy-galaxy gravitational lens system can be measured from the configuration of multiple lensed images by averaging the distances of the images from 
the center of the lensing galaxy. For two of the H-ATLAS/SDP lens candidates, 
ID81 and ID130, the positions of the lensed images are constrained by high-resolution 
SMA follow-up imaging. The lensed images of the background sources appear as peaks in the SMA signal-to-noise ratio map. Here we have selected those peaks with signal-to-noise ratio above
eight, which provided positions for four images in ID81 and two images in
ID130. The error on the Einstein radius is estimated by taking into account the
uncertainties on the position of the individual peaks. For a point source the rms error on
its position is $\sqrt{2}\sigma$/SNR (assuming no systematic astrometry errors and uncorrelated
Gaussian noise), where $\sigma$ is the Gaussian rms width of the
instrument beam ($\,$=$\,$FWHM/$2\sqrt{2{\rm ln}2}$), while SNR is the
signal-to-noise ratio at the source position ({\it 6, 7}). The SMA synthesised beam 
(derived by combining observations in VEX, COM and SUB configurations) has 
size 0.81$^{\prime\prime}\times$0.73$^{\prime\prime}$ for ID81 and 
0.78$^{\prime\prime}\times$0.72$^{\prime\prime}$ for ID130. Therefore, 
in estimating the relative positional uncertainty of the peaks, we have assumed 
FWHM=0.75$^{\prime\prime}$ and FWHM=0.77$^{\prime\prime}$ for ID81 
and ID130, respectively. The absolute positional uncertainty of the SMA images 
is estimated by referencing the data to nearby point-like sources (quasars) of 
known positions and is below 10 milli-arcseconds. \\ 
Once the Einstein radius is known, the mass within the
Einstein ring can be easily derived assuming a Singular Isothermal Sphere (SIS) model (although the result is only little dependent on the model used) 
which is characterized by a projected surface density that falls off as $\theta^{-1}$, 
where $\theta$ is the angular distance from the center of the mass distribution ({\it 8}),
\begin{equation}
  M_{\rm E} = M(<\theta_{\rm E}) = \pi\Sigma_{\rm crit}\theta^{2}_{\rm E},
\end{equation}
and $\Sigma_{\rm crit}$ is the {\it critical surface density}:
\begin{equation}
  \Sigma_{\rm crit} = \frac{c^2}{4\pi G}\frac{D_{\rm S}}{D_{\rm L}D_{\rm LS}}.
\end{equation}
In the equation above, $c$ is the speed of light, $G$ is the gravitational constant, 
$D_{\rm L}$ and $D_{\rm S}$ are the angular diameter distances to the lens and 
the source, respectively, while $D_{\rm LS}$ is the angular diameter distance 
between the lens and the source. The error on the mass is obtained by propagating 
the errors on the Einstein radius and on the spectroscopic/photometric redshifts 
used to derived the angular diameter distances. The estimated values of 
$\theta_{\rm E}$ and $M_{\rm E}$ are listed in Table$\,$2.

\clearpage
$~$ \\
{\bf Gravitational lensing modeling}
$~$ \\
\\
A detailed analysis of the lensed structure revealed by the SMA images is 
beyond the scope of this paper and is deferred to a forthcoming publication. However,
in order to prove that such a structure is consistent
with a lensing event, we have used the publicly available 
LENSMODEL software ({\it 9}) to reproduce the positions of the peaks in the SMA maps. 
We have assumed a Singular Isothermal Ellipsoid (SIE) model ({\it 8}) for
the mass distribution of the lens galaxy. The SIE model consists of concentric and aligned elliptical isodensity contours with axis ratio $q$. The circular limit is the SIS model and corresponds to $q=1$. The results are shown in Fig.$\,$S9. We have further assumed that the centroid 
of the mass model coincides with that of the light distribution of the lensing galaxy.
The best-fit model for ID130 has ellipticity $e=0.16$ and position angle (measured East of North) of $\theta=+75\,$deg, consistent with the results found for the optical light-distribution that is dominated by the more compact Sersic profile (Table$\,$S4 and Fig.$\,$S8). For ID81, the mass distribution has ellipticity $e=0.24$ and position angle $\theta=-3\,$deg, which is not consistent with that measured for the luminous component (Table$\,$S4 and Fig.$\,$S8). Besides, the position of the peaks is not well reproduced by the model. This may hint at the effect of an external shear (which we did not include) due to a nearby cluster (photometrically detected 3.6 arcminutes away), in the direction indicated by the arrow in Fig.$\,$S9. \\
We have used the best-fit lens models to approximatively quantify the magnification experienced by a background source described by a Gaussian profile with a Full Width at Half Maximum (FWHM) in the range 0.1-0.3$^{\prime\prime}$. This extension is consistent with the physical size of the submillimeter galaxy studied by ({\it 10}). The inferred magnification is $\sim$18-31 for ID81 and $\sim$5-7 for ID130.
An example of lensed image, after convolution with the SMA point spread function, for the case FWHM=0.2$^{\prime\prime}$ is shown in Fig.$\,$S9.

\clearpage

$~$ \\
{\bf {\LARGE Figures}} \\
$~$ \\

\setcounter{figure}{0}\renewcommand{\figurename}{Figure S}
\begin{figure}[!bht]\caption{CO line detections in ID81 and ID130. The figure
shows the difference between the spectrum of ID81 and that of ID130,
derived from Zpectrometer observations. The relative spectrum is
normalized such that the peak line strength of ID81 is equal to 1. In
both objects the peak is associated with the CO J=1-0 emission line
(rest-frame frequency 115.27$\,$GHz).}
  \hspace{+2.0cm}\includegraphics[height=8.5cm,width=12cm]{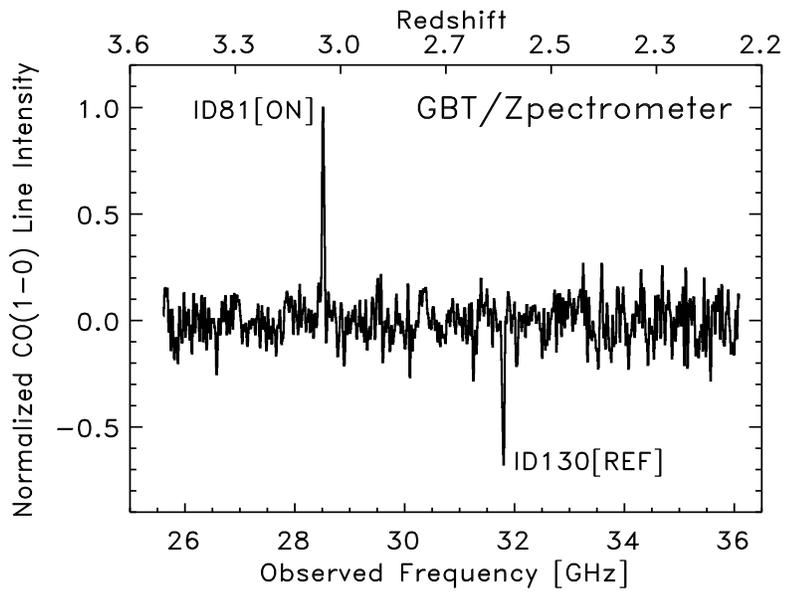} \\
  \vskip-0.5truecm
\end{figure}

\setcounter{figure}{1}\renewcommand{\figurename}{Figure S}
\begin{figure}\caption{CO line detections in ID9. The figure shows the
spectrum derived from Z-Spec observations. The CO emission lines
redshifted into the frequency range probed by Z-Spec correspond to
transitions J=5-4 (rest-frame frequency 576.3$\,$GHz) and J=6-5
(rest-frame frequency 691.5$\,$GHz).}
  \hspace{+0.0cm}\includegraphics[angle=90,height=6.0cm,width=16cm]{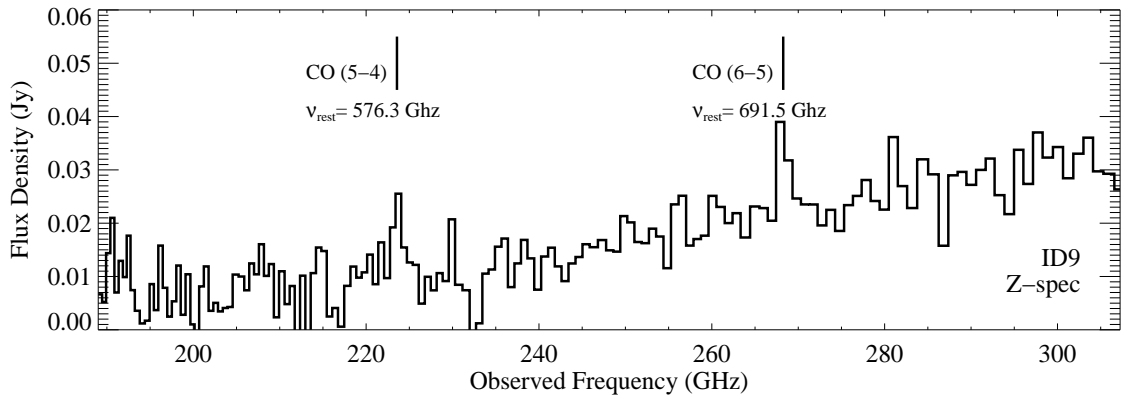} \\
  \vskip-0.5truecm
\end{figure}

\setcounter{figure}{2}\renewcommand{\figurename}{Figure S}
\begin{figure}\caption{Optical spectra of ID11 and ID17 obatined with the WHT.}
  \vspace{-3.0cm}
  \hspace{-1.0cm}
  \begin{minipage}[b]{0.5\linewidth}
    \centering \resizebox{1.2\hsize}{!}{
      \hspace{0.0cm}\includegraphics{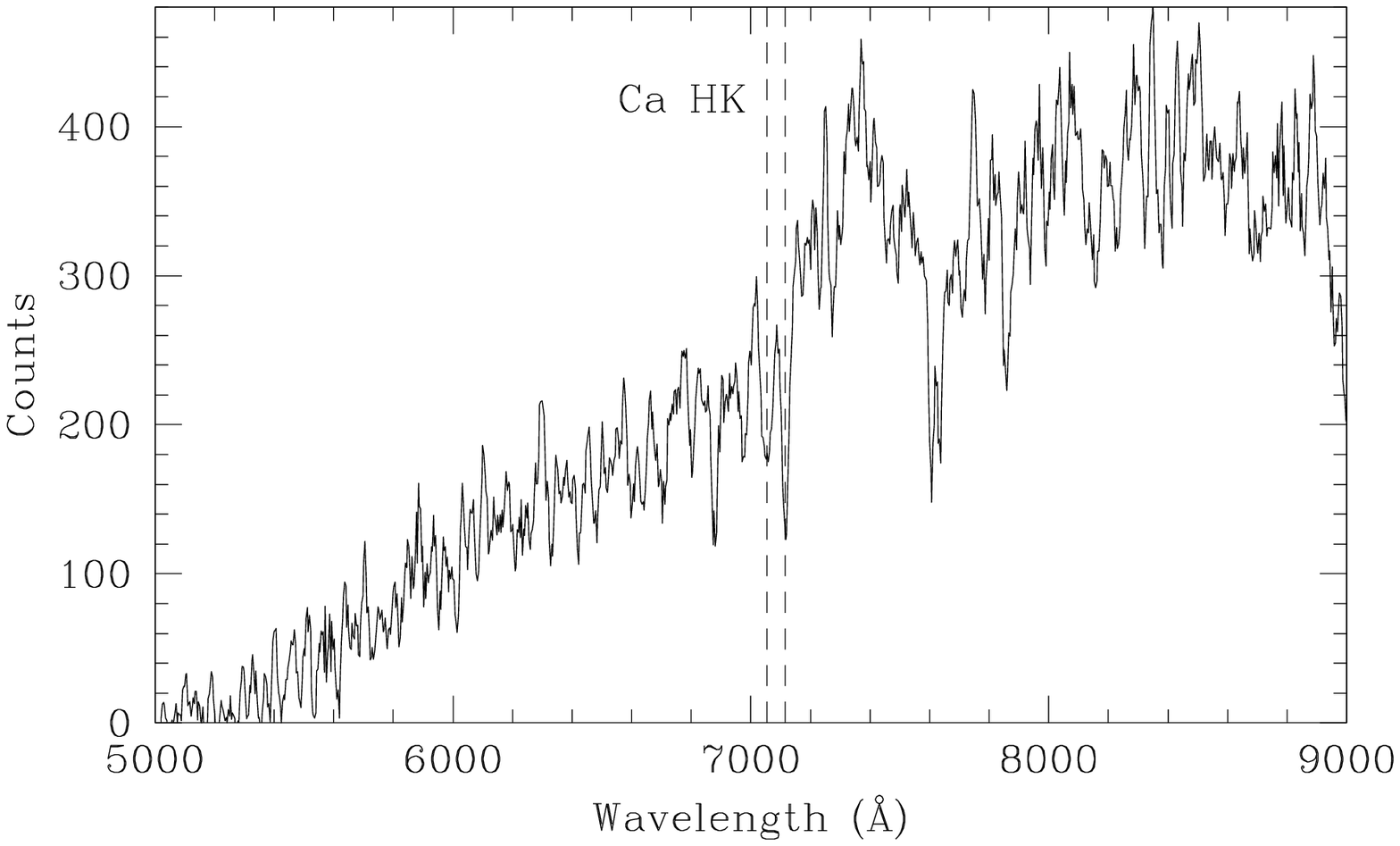}
}
  \end{minipage}
  \hspace{+0.5cm}
  \begin{minipage}[b]{0.5\linewidth}
    \centering \resizebox{1.2\hsize}{!}{
      \hspace{0.0cm}\includegraphics{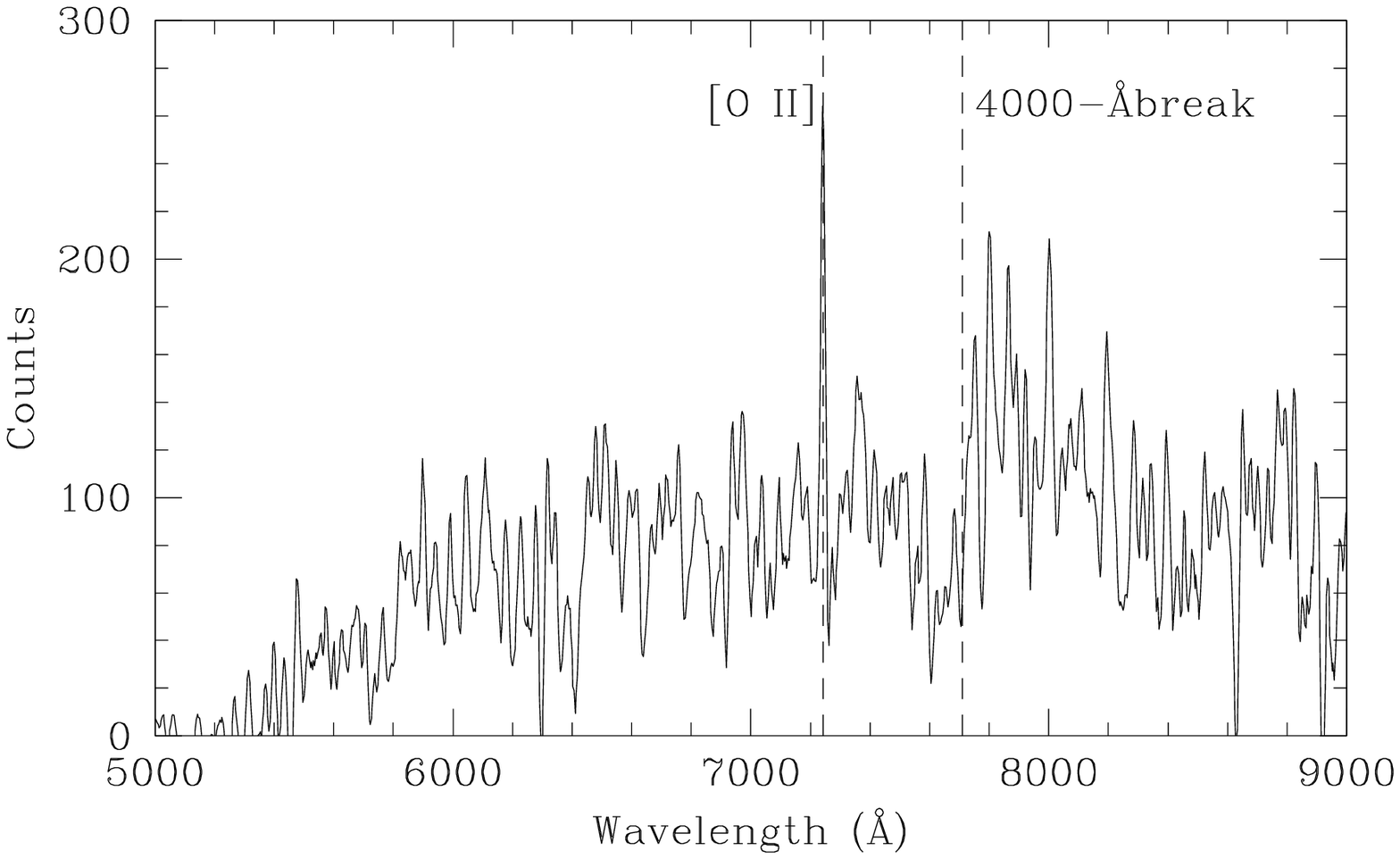}
 }
  \end{minipage}
\end{figure}

\setcounter{figure}{3}\renewcommand{\figurename}{Figure S}
\begin{figure}\caption{Optical spectrum of ID130 obtained with the APO
telescope. The bottom panel of the figure shows the reduced 2-d
spectrum in the region of the detected emission lines.  The top panel
shows the flux summed in a 5-pixel wide (2 arcsecond) aperture
centered on the object, with an arbitrary flux scale because the
clouds made wavelength calibration impossible.}
  \begin{center}
    \hspace{-1.2cm}\includegraphics[width=1.0\textwidth]{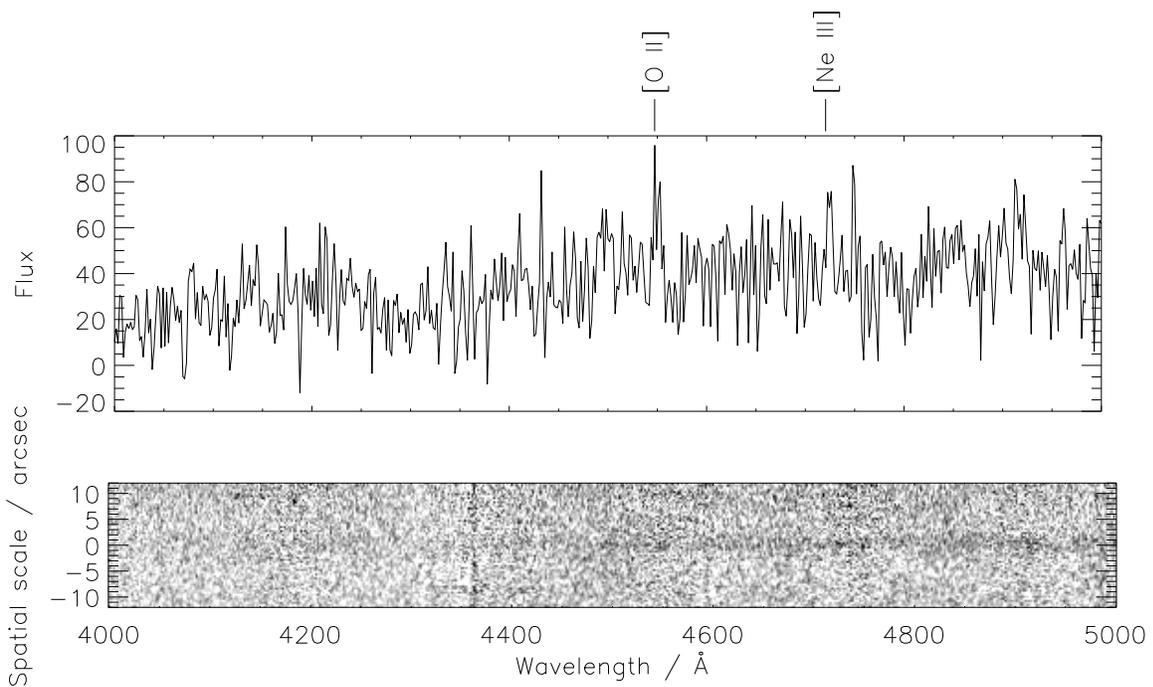}
  \end{center}
\end{figure}

\setcounter{figure}{4}\renewcommand{\figurename}{Figure S}
\begin{figure}\caption{Best-fit to the light distribution of the lens galaxy
in the gravitational lens systems ID9, ID11 and ID17. The postage
stamp images show, from left to right, the keck i-band image, the
best-fit light distribution model provided by GALFIT and the residual
map obtained by subtracting the best-fit model from the observed light
distribution. The map of the residuals show no evident structure,
implying that the background source is particularly faint in the
optical, despite the magnification due to lensing.}
  \begin{center}
    \includegraphics[width=0.99\textwidth]{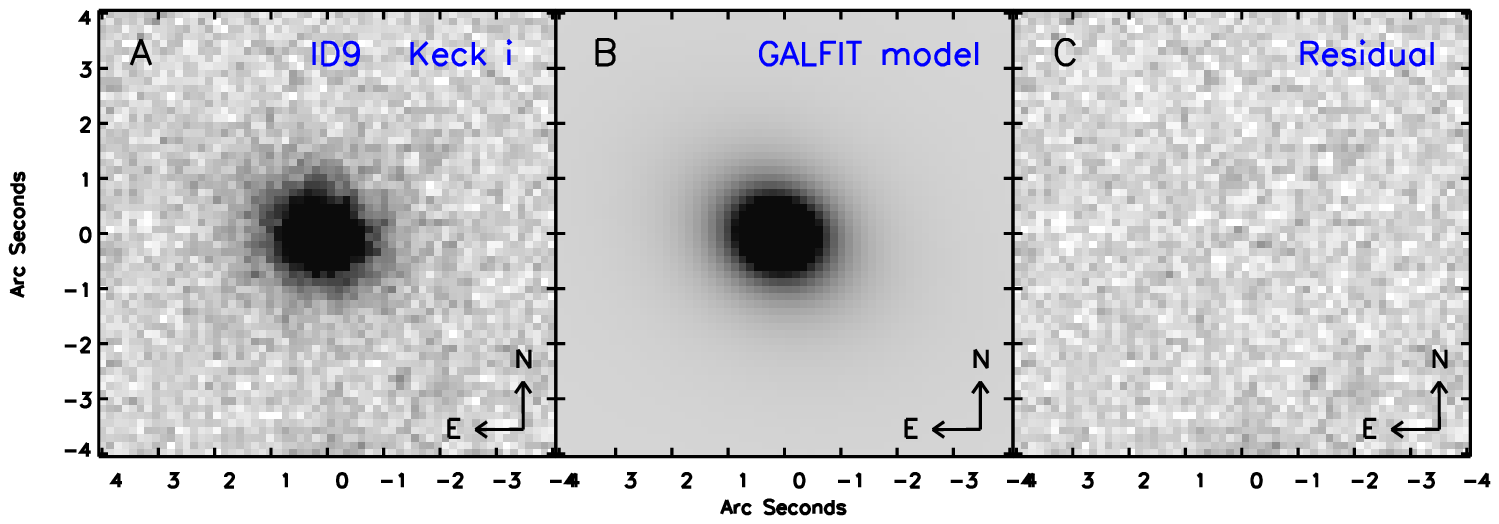}
    \includegraphics[width=0.99\textwidth]{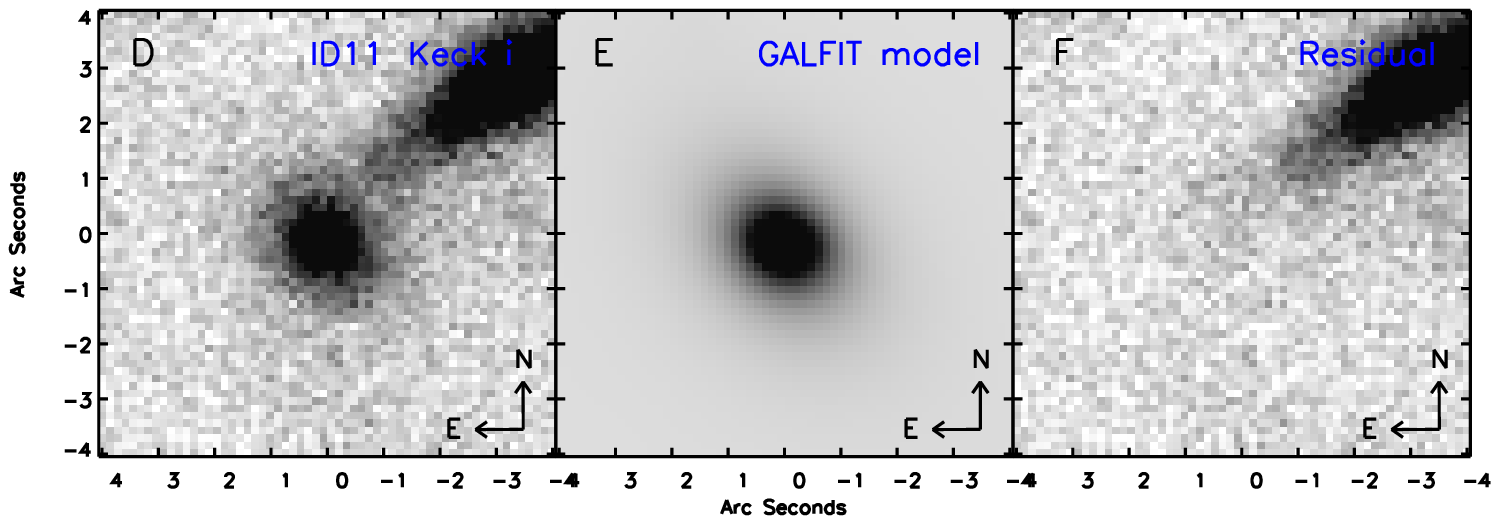}
    \includegraphics[width=0.99\textwidth]{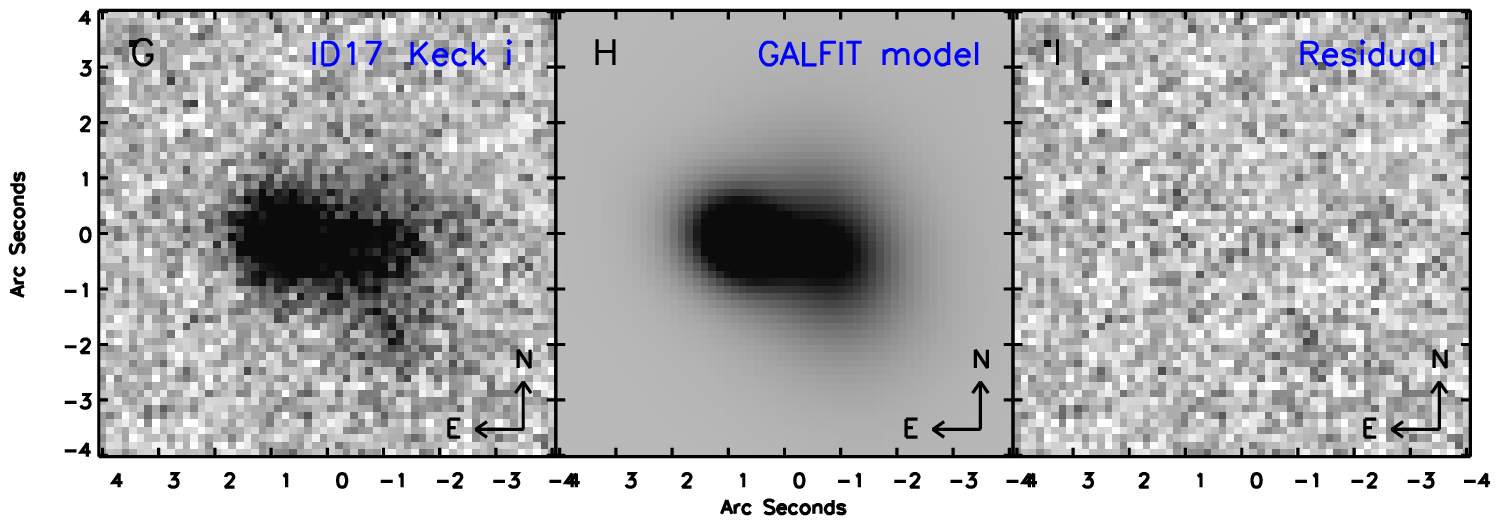}
  \end{center}
\end{figure}

\setcounter{figure}{5}\renewcommand{\figurename}{Figure S}
\begin{figure}\caption{Best-fit to the light distribution of the lens galaxy
in the gravitational lens systems ID81 and ID130. The postage stamp
images show, from left to right, the keck i-band image, the best-fit
light distribution model provided by GALFIT and the residual map
obtained by subtracting the best-fit model from the observed light
distribution. The SMA contours (in red) are overlaid on the optical
images (in steps of 6$\sigma$, 8$\sigma$, 10$\sigma$, etc.) to
highlight that there is no evident correspondence between the submillimeter
and the optical emission in the residual maps.}
  \begin{center}
     \includegraphics[width=0.99\textwidth]{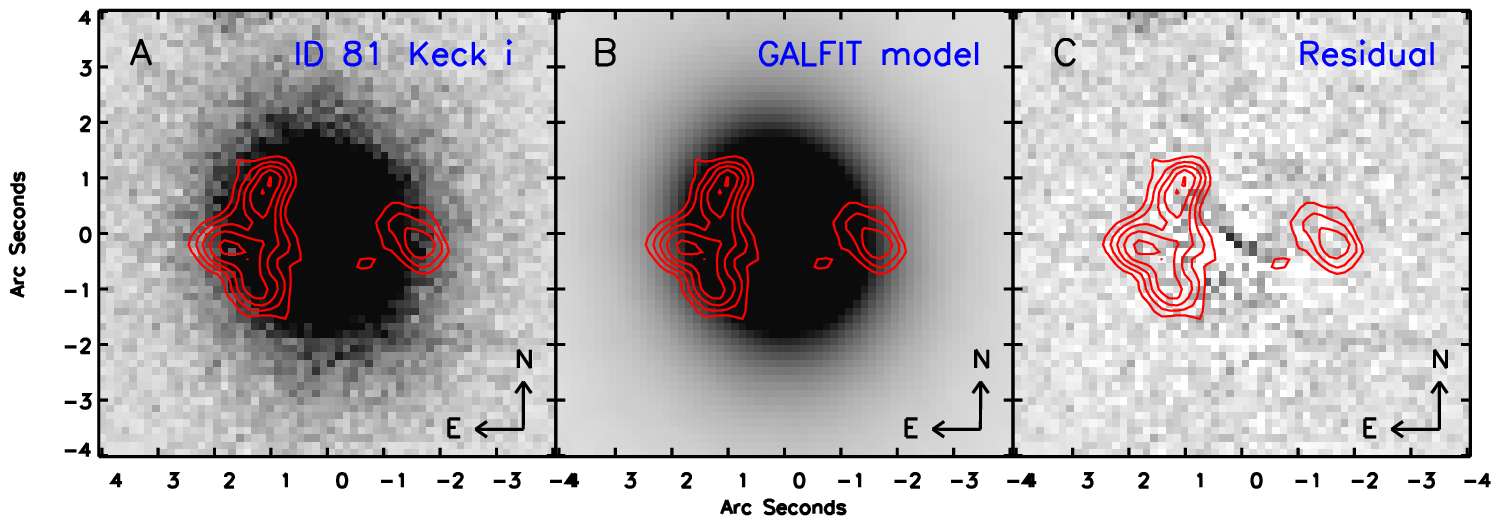}
     \includegraphics[width=0.99\textwidth]{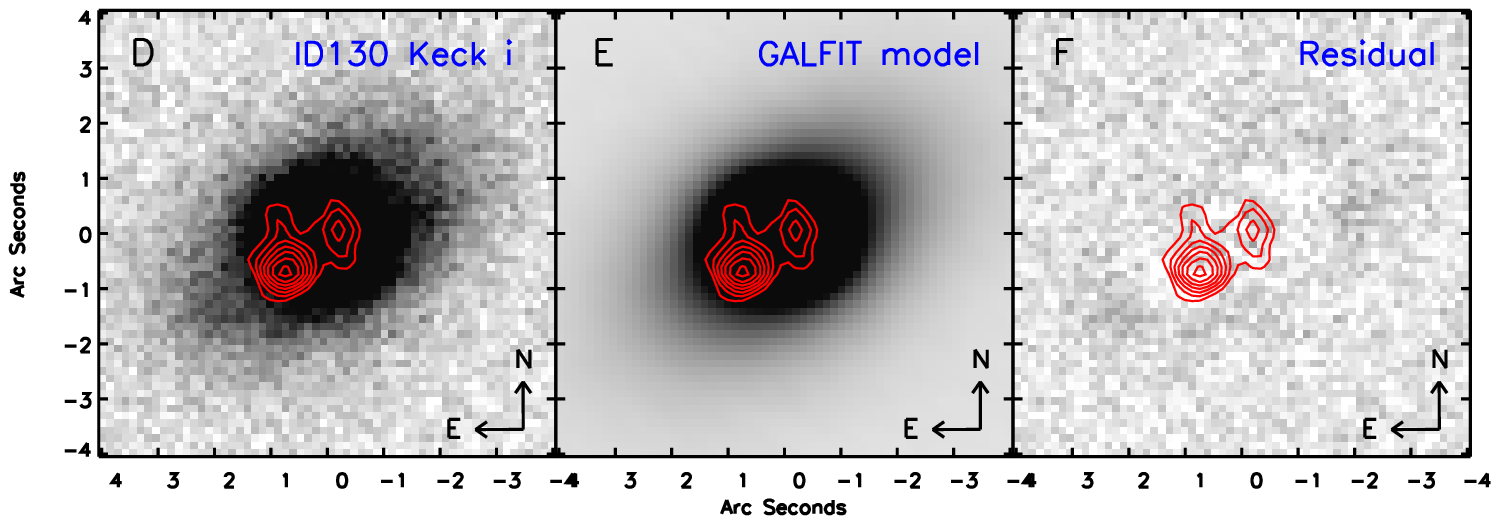}
  \end{center}
\end{figure}

\setcounter{figure}{6}\renewcommand{\figurename}{Figure S}
\begin{figure}\caption{Decomposition of the best-fit models of the lens
galaxies in ID17, ID81 and ID130. {\it Top}: ID17 shows two partially
superimposed components, indicative of two distinct lens galaxies,
each described by a relatively shallow Sersic profile. {\it Middle}:
ID81 has one single lens galaxy whose light profile is reproduced by
the sum of a Sersic profile and an exponential disk profile. {\it
  Bottom}: ID130 is similar to ID81, with the light profile being
described by the superposition of a compact Sersic profile and an
exponential disk profile. In both ID81 and ID130, the SMA contours (in
red) are overlaid on the optical images, in steps of 6$\sigma$,
8$\sigma$, 10$\sigma$, etc.}
  \begin{center}
    \includegraphics[width=0.7\textwidth]{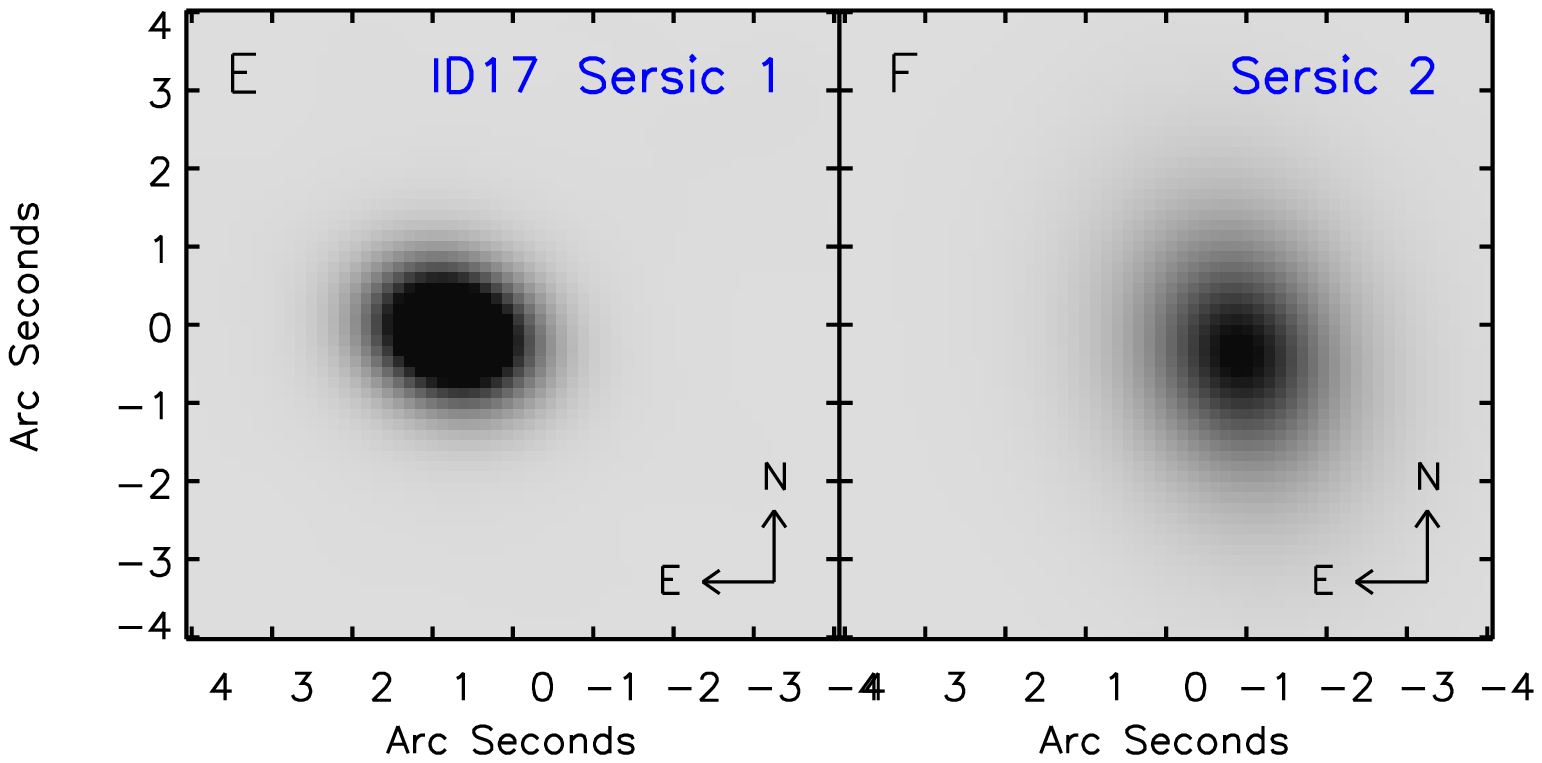}
    \includegraphics[width=0.7\textwidth]{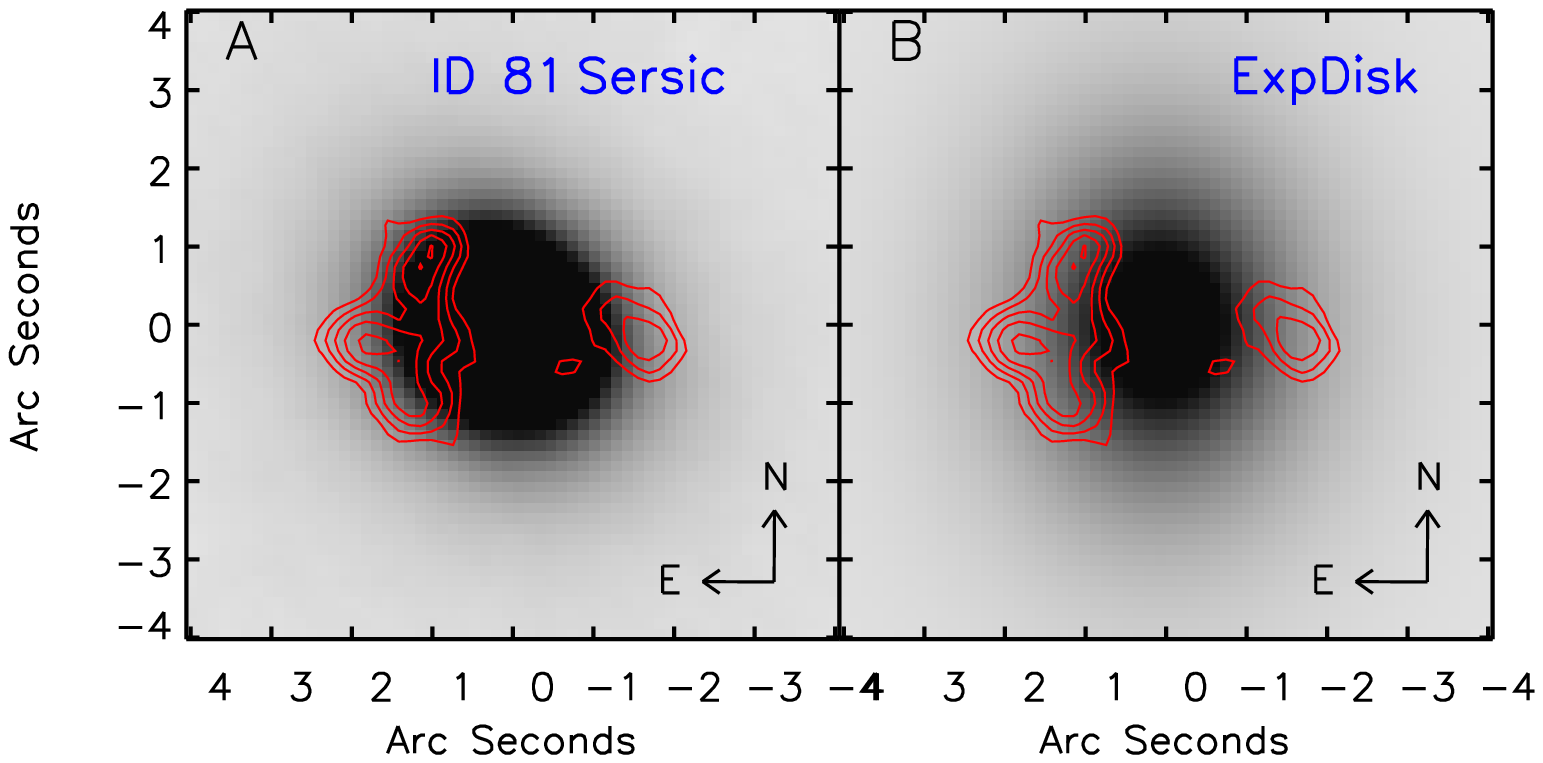}
    \includegraphics[width=0.7\textwidth]{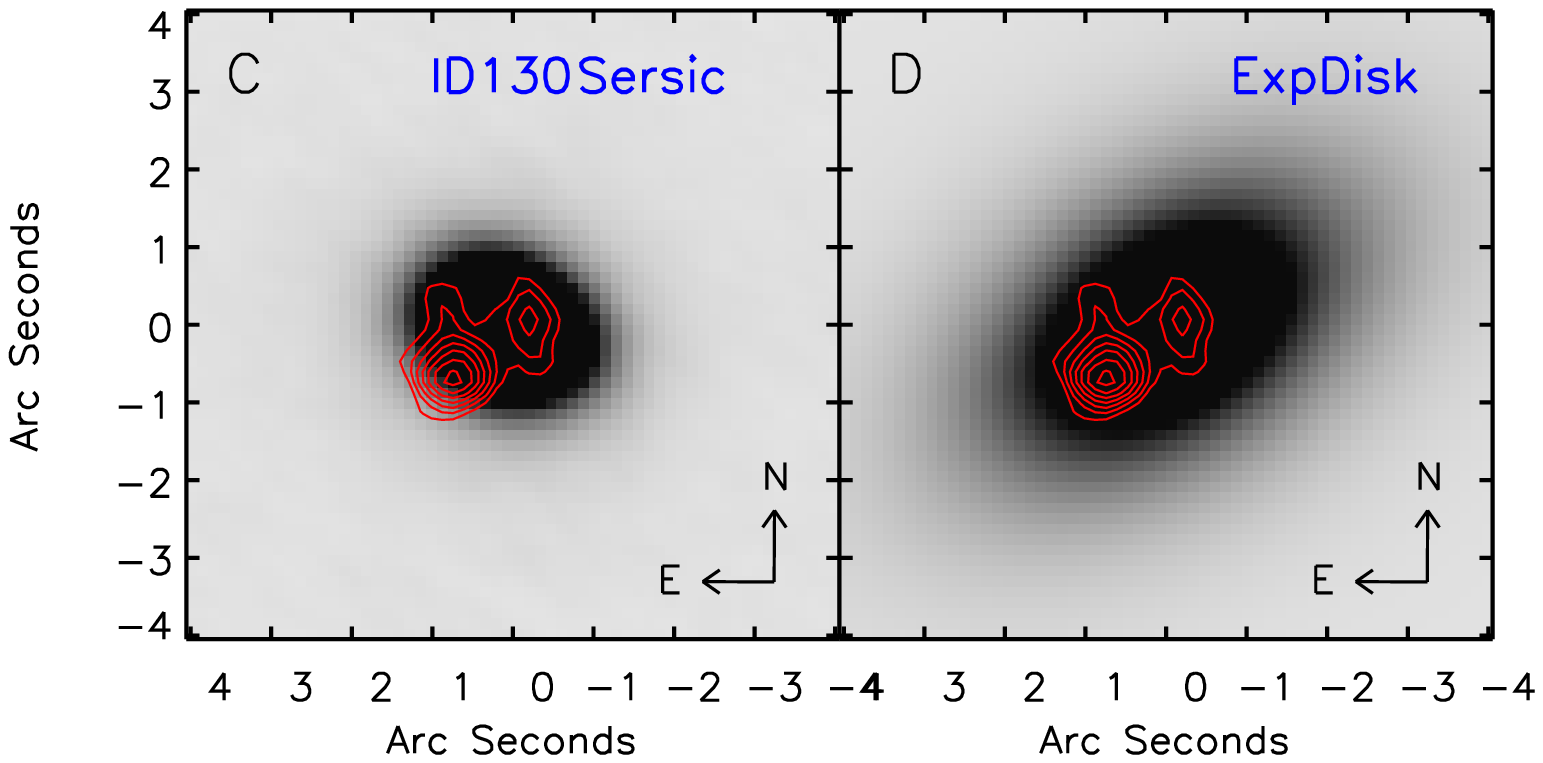}
  \end{center}
\end{figure}

\setcounter{figure}{7}\renewcommand{\figurename}{Figure S}
\begin{figure}\caption{Light profile of the lens galaxy in the gravitational
lens systems ID81 and ID130. In both cases the best-fit to the
observed light distribution of the lens galaxy is achieved using 2
components, i.e. an inner (more compact) Sersic profile and an
exponential disk profile. These components are shown as a function of
the distance from the galaxy center for ID81 (left-hand panel) and
ID130 (right-hand panel).}
  \hspace{-0.5cm}
  \begin{minipage}[b]{0.5\linewidth}
    \centering \resizebox{1.0\hsize}{!}{
      \hspace{0.0cm}\includegraphics{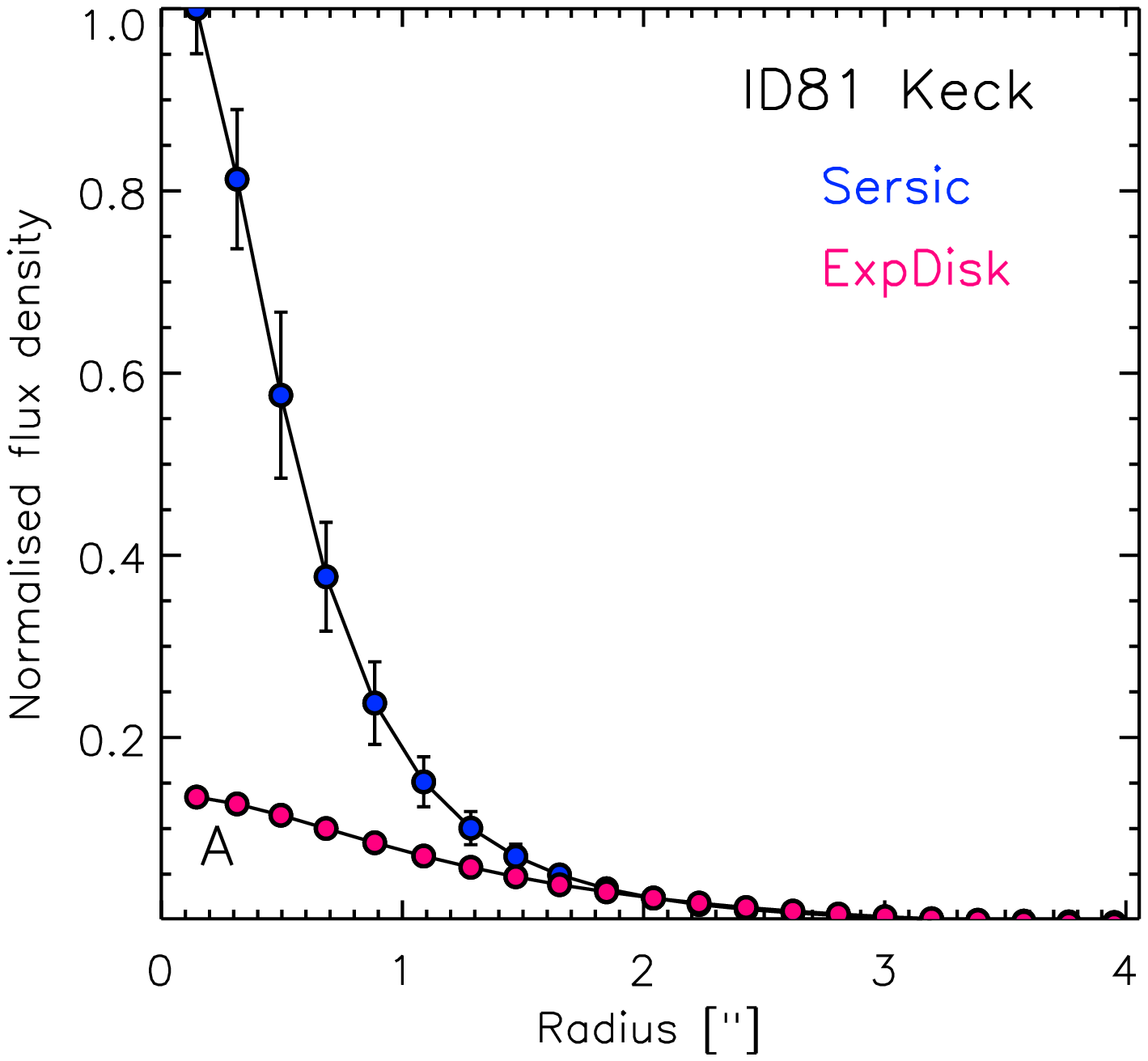}
}
  \end{minipage}
  \hspace{-0.5cm}
  \begin{minipage}[b]{0.5\linewidth}
    \centering \resizebox{1.0\hsize}{!}{
      \hspace{0.0cm}\includegraphics{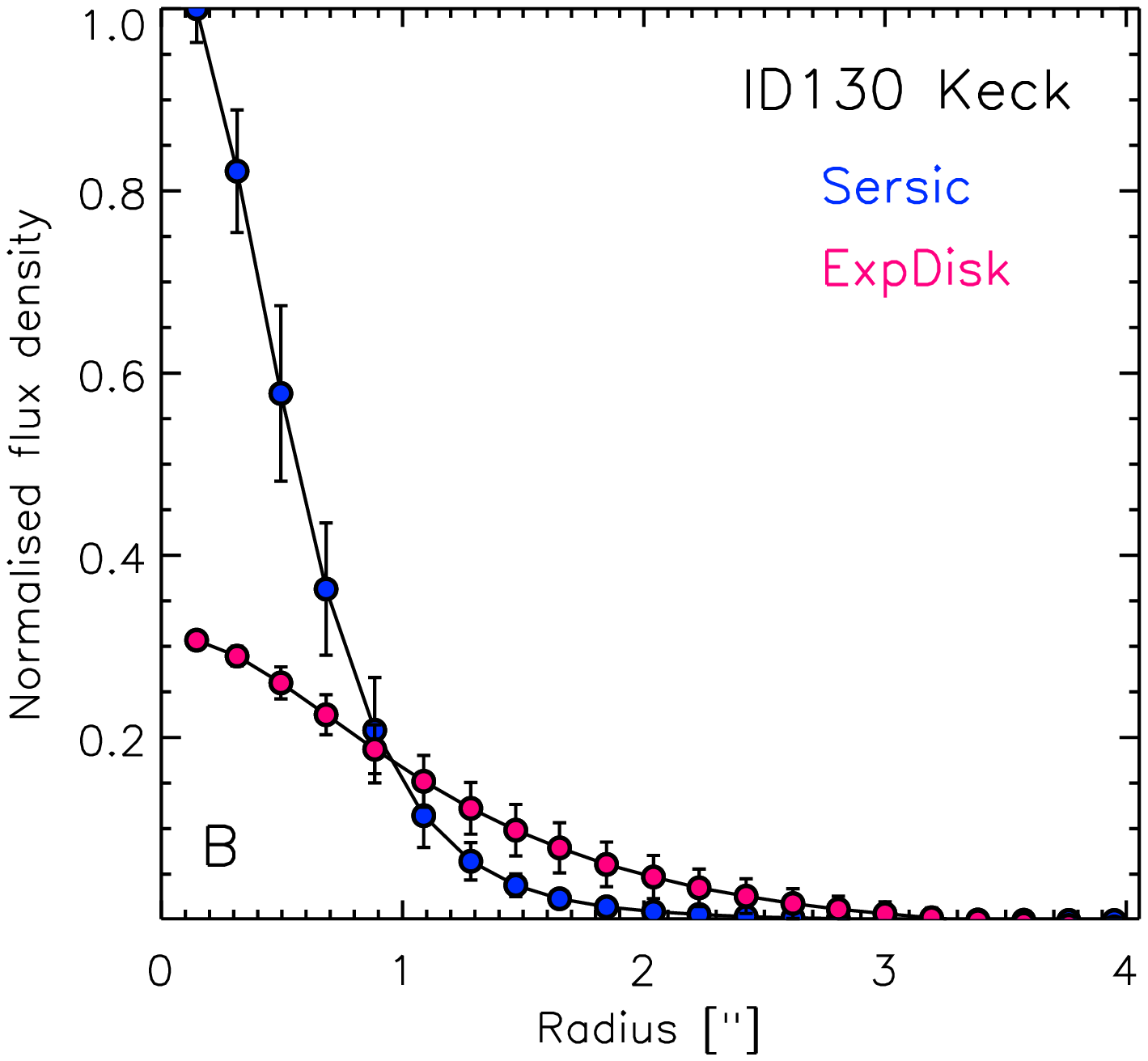}
 }
  \end{minipage}
\end{figure}

\setcounter{figure}{8}\renewcommand{\figurename}{Figure S}
\begin{figure}\caption{Lens modelling of ID81 and ID130. The LENSMODEL
software was used to fit the position of the peaks in the SMA
images. A SIE model was assumed for the mass distribution of the
foreground lenses. The image positions used in the fit are indicated
by the red dots and correspond to the peaks in the SMA images (red
contours in steps of 6$\sigma$, 8$\sigma$, 10$\sigma$ etc.). The blue
stars are the best-fit positions of the lensed background source,
assumed to be point-like. The best-fit position of the lens galaxy and
of the background source are marked by the yellow and the blue dots,
respectively. The caustic lines and critical lines of the best-fit
lens model are indicated in green and yellow, respectively, while the
yellow dashed line shows the major axis of the mass model. The
simulated image shows the lensed image of a background source (after
convolution with the SMA point spread function and added noise)
described by a Gaussian profile with FWHM$=0.2^{\prime\prime}$.}
  \vspace{+1.0cm}
  \hspace{-3.5cm}
  \begin{minipage}[b]{0.5\linewidth}
    \centering \resizebox{1.5\hsize}{!}{
      \hspace{0.0cm}\includegraphics{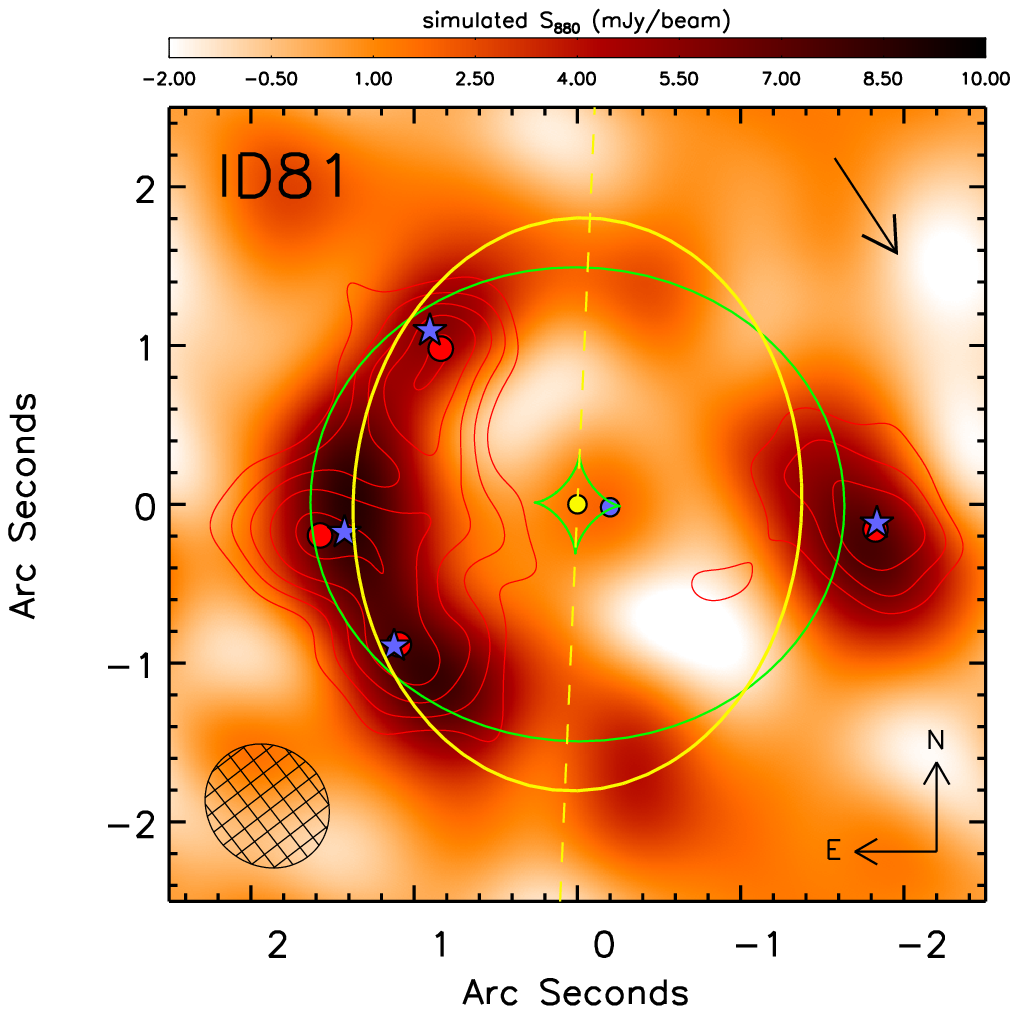}
}
  \end{minipage}
  \hspace{-0.5cm}
  \begin{minipage}[b]{0.5\linewidth}
    \centering \resizebox{1.5\hsize}{!}{
      \hspace{0.0cm}\includegraphics{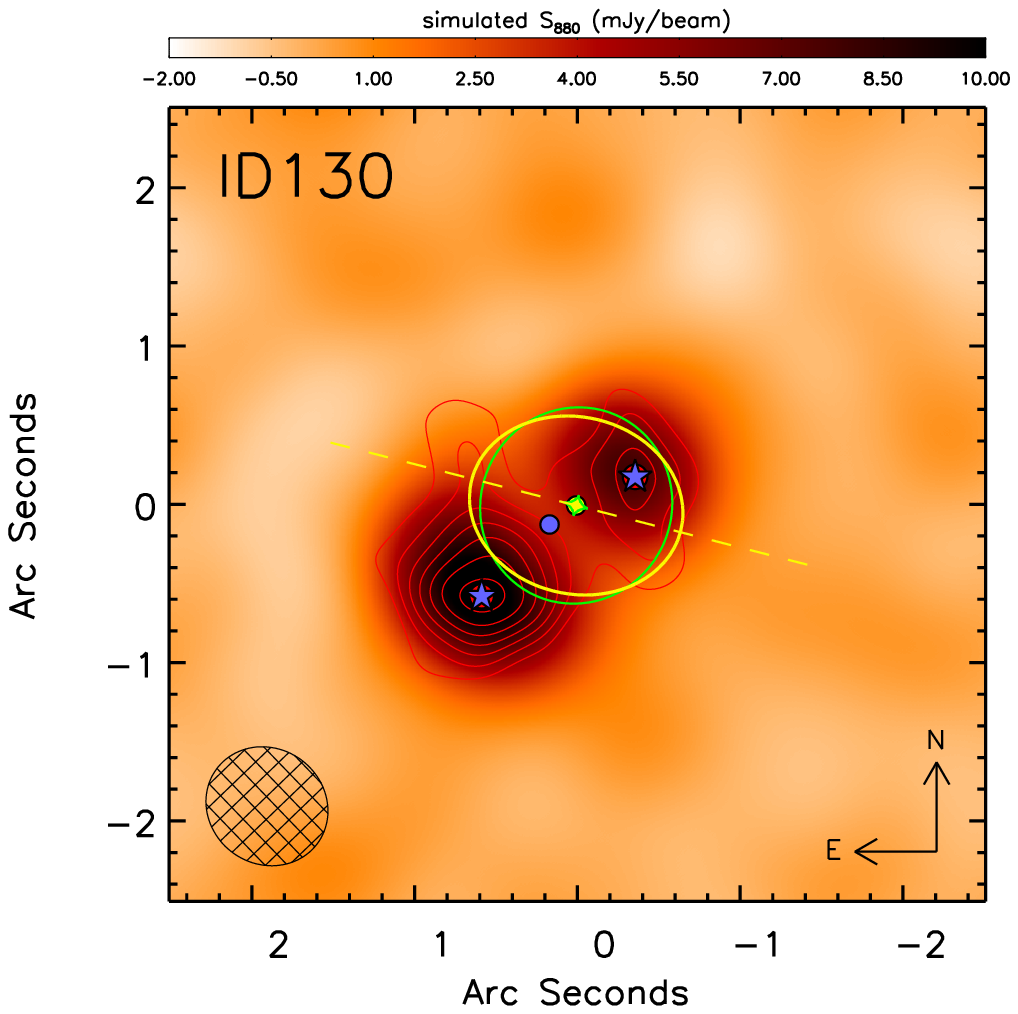}
 }
  \end{minipage}
\end{figure}

\clearpage

$~$ \\
{\bf {\LARGE Tables}} \\
$~$ \\

\setcounter{table}{0}\renewcommand{\tablename}{Table S}
\begin{table*}[!bht]
  \vspace{0.0cm}
  \begin{center} \caption{Submillimeter/millimeter fluxes for the lens candidates. The
official H-ATLAS ID, according to IAU (International Astronomical
Union) convention, is that derived from the position of the sources in
the H-ATLAS SDP catalogue. The quoted errors on the {\it Herschel}
flux densities include statistical errors, confusion noise and an
absolute calibration uncertainty of 10 per cent for PACS/100$\mu$m, 20
per cent for PACS/160$\mu$m and 15 per cent for SPIRE. 3$\sigma$ upper
limits at PACS wavelengths are provided when no detection is achieved
at that significance level. Note that ID130 lies just outside the
region covered by PACS. Fluxes at 880$\,\mu$m are from the SMA, while
those at 1200$\,\mu$m are from the MAMBO array at the Institut de
Radioastronomie Millim\'etrique (IRAM) 30$\,$m telescope.}
    \scriptsize
    \vspace{+0.5cm}
    \begin{tabular}{ccrrrrrrr}
      \hline 
      \multicolumn{1}{c}{H-ATLAS ID} 
      & \multicolumn{1}{c}{SDP ID}
      & \multicolumn{1}{c}{$S_{\rm100}$} 
      & \multicolumn{1}{c}{$S_{\rm160}$} 
      & \multicolumn{1}{c}{$S_{\rm250}$} 
      & \multicolumn{1}{c}{$S_{\rm350}$} 
      & \multicolumn{1}{c}{$S_{\rm500}$} 
      & \multicolumn{1}{c}{$S_{\rm880}$} 
      & \multicolumn{1}{c}{$S_{\rm1200}$} 
      \\
      &  \multicolumn{1}{c}{~}  
      &  \multicolumn{1}{c}{(mJy)} 
      &  \multicolumn{1}{c}{(mJy)}  
      &  \multicolumn{1}{c}{(mJy)}  
      &  \multicolumn{1}{c}{(mJy)} 
      &  \multicolumn{1}{c}{(mJy)} 
      &  \multicolumn{1}{c}{(mJy)}
      &  \multicolumn{1}{c}{(mJy)} \\
      \hline
      \hline
      H-ATLAS J090740.0$-$004200  &  9    &  187$\pm$57  &  416$\pm$94  &  485$\pm$73  &  323$\pm$49  &  175$\pm$28  &  $-$             &  7.6$\pm$0.8 \\
      H-ATLAS J091043.1$-$000321  &  11   &  198$\pm$55  &  397$\pm$90  &  442$\pm$67  &  363$\pm$55  &  238$\pm$37  &  $-$             &  12.2$\pm$1.2  \\
      H-ATLAS J090302.9$-$014127  &  17   &  78$\pm$55  &  182$\pm$56  &  328$\pm$50  &  308$\pm$47  &  220$\pm$34  &  $-$             &  15.3$\pm$1.3 \\
      H-ATLAS J090311.6+003906  &  81   &  $\le$62     &  $\le$83     &  129$\pm$20  &  182$\pm$28  &  166$\pm$27  &  76.4$\pm$3.8  &  20.0$\pm$0.7  \\
      H-ATLAS J091305.0$-$005343  &  130  &        $-$     &        $-$     &  105$\pm$17  &  128$\pm$20  &  108$\pm$18  &  39.3$\pm$2.3  &  11.2$\pm$1.2  \\
      \hline
    \end{tabular} 
  \end{center}
\end{table*}

\setcounter{table}{1}\renewcommand{\tablename}{Table S}
\begin{table*}[!bht]\caption{UV/optical/NIR photometry for the lens candidates. UV
data are from GALEX, optical photometry is from SDSS and NIR data are
from UKIDSS [as reprocessed by GAMA; ({\it 11})]. Where photometric
measurements are not listed it means that the source is not covered at
those wavelengths. The 3-$\sigma$ upper limits shown within
parenthesis for the UKIDSS wavelengths are obtained from the residual
image after the best-fit GALFIT model of the source has been
subtracted off. These limits are used to constrain the SED of the
background source.}
  \vspace{0.0cm}
  \begin{center}
    \footnotesize
    \begin{tabular}{c|ccccc}
      \hline 
      \multicolumn{1}{c|}{SDP ID}
      & \multicolumn{1}{|c}{9} 
      & \multicolumn{1}{c}{11}
      & \multicolumn{1}{c}{17} 
      & \multicolumn{1}{c}{81} 
      & \multicolumn{1}{c}{130} \\
      \hline
      \hline
      GALEX FUV ($\mu$Jy) & -                  & -                    & -                    & 0.23$\pm$0.18       &  - \\
      GALEX NUV ($\mu$Jy) & -                  & -                    & -                    & 1.9$\pm$1.1         & -  \\
      SDSS u ($\mu$Jy)    & 0.24$\pm$0.23      & 0.57$\pm$0.59        & 3.3$\pm$1.6          & 3.9$\pm$2.0         & 1.7$\pm$1.7  \\
      SDSS g ($\mu$Jy)    & 1.79$\pm$0.43      & 1.01$\pm$0.45        & 3.9$\pm$6.4          & 24.9$\pm$1.1        & 19.41$\pm$0.72  \\
      SDSS r ($\mu$Jy)    & 5.81$\pm$0.70      & 3.94$\pm$0.65        & 7.7$\pm$1.0          & 114.8$\pm$2.1       & 66.1$\pm$1.2  \\
      SDSS i ($\mu$Jy)    & 14.9$\pm$1.1       & 11.3$\pm$1.0         & 15.3$\pm$1.5         & 197.7$\pm$3.6       & 108.6$\pm$2.0  \\
      SDSS z ($\mu$Jy)    & 27.0$\pm$3.7       & 21.5$\pm$4.2         & 11.8$\pm$6.0         & 278.0$\pm$3.6       & 143.2$\pm$6.6  \\
      UKIDSS Y ($\mu$Jy)  & -                  & -                    & 27.7$\pm$9.5($<$6.6) & 321.3$\pm$3.2($<$6.3) & - \\
      UKIDSS J ($\mu$Jy)  & -                  & 102.4$\pm$9.8($<$16) & 56$\pm$17($<$12)     & 367$\pm$11($<$9.2)    & - \\
      UKIDSS H ($\mu$Jy)  & 73$\pm$15($<$5.0)  & 237$\pm$17($<$14)    & 107$\pm$19($<$8.2)   & 508.1$\pm$5.3($<$8.5) & - \\
      UKIDSS K ($\mu$Jy)  & 132$\pm$24($<$6.5) & -                    & 108$\pm$23($<$9.7)   & 573.7$\pm$6.2($<$14)  & - \\
      \hline
    \end{tabular} 
  \end{center}
\end{table*}

\setcounter{table}{2}\renewcommand{\tablename}{Table S}
\begin{table*}\caption{Technical information on the SMA follow-up
observations. This includes: the date the measurements were taken
(Date), the configuration of the antennas (Conf.; VEX=`very-extended',
SUB=`sub-compact', COM=`compact'), the number of antennas used
(Nant.), the projected baselines lengths (min/mean/max Pr Baselines),
the Local Oscillator Frequency (LO Freq.), and the on$-$source
integration time (Int. time).}
  \vspace{0.0cm}
  \begin{center}
    \begin{minipage}{15cm}
      \normalsize
      \vspace{+0.5cm}
      \begin{tabular}{ccccccc}
        \hline
        \multicolumn{1}{c}{SDP ID} 
        & \multicolumn{1}{c}{Date}
        & \multicolumn{1}{c}{Conf.}
        & \multicolumn{1}{c}{Nant.} 
        & \multicolumn{1}{c}{min/mean/max}
        & \multicolumn{1}{c}{LO Freq.}
        & \multicolumn{1}{c}{Int. time}
        \\
      
        & 
        & 
        &
        & \multicolumn{1}{c}{Pr Baselines (m)}
        & \multicolumn{1}{c}{(GHz)\footnote{\scriptsize{Total bandwidth coverage is LO-8 to LO-4 (LSB) and LO+4 to LO+8 (USB)
              for a total of ~8 GHz.  The small difference in LO Frequency between compact
              configuration observations and the subcompact and very extended
              observations is not important in this context.}} }   
        & \multicolumn{1}{c}{(min)}
        \\
        \hline
        \hline
        81 & 25Feb10  &  VEX  &  7  &  69/281/509  &  340.7  &   289 \\
        130 & 28Feb10  & VEX  &  7  &  76/289/509  &  340.7  &     298 \\
        81 & 16Mar10 &  SUB &   5  &   6/ 17/ 25  &  340.7 &      144 \\
        130 &      &      &              &        &    &    152 \\
        81 & 09Apr10 &  COM &   6   &  9/ 38/ 69  &  341.6 &      153 \\
        130 & &       &      &             &             &    144 \\
        \hline
      \end{tabular} 
    \end{minipage}
  \end{center}
\end{table*}

\setcounter{table}{3}\renewcommand{\tablename}{Table S}
\begin{table*}[!htb]\caption{GALFIT results for the five gravitational lens
systems. The 3$-\sigma$ upper limits given are for an extended source
and derived from the distribution of 1.5 arcsecond radius aperture
photometry of the Keck maps.}
  \vspace{0.0cm}
  \begin{center}
  \begin{minipage}{15cm}
    \begin{tabular}{@{}cccccccccc}
      \hline
H-ATLAS ID & profile\footnote{ExpDisk = exponential disk profile} & $\chi^2_\nu\footnote{Reduced $\chi^2$}$ & radius\footnote{Radius for Sersic and disk scale length for ExpDisk} & $\delta$\footnote{Sersic index} &  Axis ratio & Angle\footnote{Angle measured east of north }& g-3$\sigma$ & i-3$\sigma$\\
      & &  & (arcsec) &  &    & deg. &  ($\mu$Jy) & ($\mu$Jy) \\
      \hline
       \hline
       9 & Sersic & 1.07  & 0.85 & 5.36 & 0.72 & 56.76 & 0.162 & 0.641 \\
      11 & Sersic & 1.03 & 1.10 & 2.97 & 0.65 & 39.61 & 0.229 & 0.442  \\
      17 & Sersic & 1.07 & 0.61 & 0.54 & 0.71 & 63.25 & 0.202 & 0.404  \\
      & Sersic & $-$ & 1.36 & 0.91 & 0.69 & 12.83 & $-$ & $-$ \\
      81 & Sersic & 1.13 & 0.70 & 2.82 & 0.78 & 36.45 & 0.130 & 0.202 \\ 
      & ExpDisk & $-$ & 1.20 & $-$ & 0.72 & 0.62 & $-$ & $-$\\
      130 & Sersic & 1.00 & 0.32 & 1.23 &  0.52 & 56.82 & 0.198 & 0.351 \\
       & ExpDisk & $-$ & 1.11 & $-$ & 0.55 & -54.64 & $-$ & $-$ \\
      \hline
    \end{tabular}
  \end{minipage}
  \end{center}
\end{table*}

\clearpage

\subsection*{References and Notes}
\begin{itemize}
\item[1.]
J. B. Oke, {\it et al.}, {\it Publ. Astron. Soc. Pac.} {\bf 107}, 375 (1995).
\item[2.]
J. K. McCarthy, {\it et al.}, {\it Society of Photo-Optical Instrumentation Engineers (SPIE) Conference}, S. D'Odorico, ed. (1998), vol. 3355, pp. 81-92
\item[3.]
S. Guilloteau, {\it et al.}, {\it Astron. \& Astrophys.} {\bf 262}, 624 (1992)
\item[4.]
http://www.apo.nmsu.edu/arc35m/Instruments/DIS/
\item[5.]
C. Y. Peng, L. C. Ho, C. D. Impey, H. Rix, {\it Astron. J.} {\bf 124}, 266 (2002)
\item[6.]
J. J. Condon, {\it Publ. Astron. Soc. Pac.} {\bf 109}, 166 (1997)
\item[7.]
R. J. Ivison, {\it et al.}, {\it Mon. Not. R. Astron. Soc.} {\bf 380}, 199 (2007)
\item[8.]
R. Kormann, P. Schneider,  M. Bartelmann, {\it Astron. \& Astrophys.} {\bf 284}, 285 (1994)
\item[9.]
C. R. Keeton, {\it available at http://xxx.lanl.gov/abs/astro-ph/0102340 (2001)}
\item[10.]
A. M. Swinbank, {\it et al.}, {\it Nature} {\bf 464}, 733 (2010)
\item[11.]
D. Hill, {\it et al.}, {\it available at http://xxx.lanl.gov/abs/astro-ph/1009.0615}
\end{itemize}

\end{document}